\documentclass[12pt]{elsarticle}
\usepackage[utf8]{inputenc}
\usepackage{graphicx}
\usepackage{subfigure}

\usepackage{amssymb}
\usepackage{amsthm}
\usepackage{url}
\usepackage{amsmath}
\usepackage{subfigure}
\usepackage{textcomp}
\usepackage{xcolor}

\graphicspath{{figs/}}

\journal{}

\begin{document}

\begin{frontmatter}

\title{On Boolean gates in fungal colony}
\author[1,*]{Andrew Adamatzky}
\author[2]{Martin Tegelaar}
\author[2]{Han A. B. Wosten} 
\author[1]{\\Anna L. Powell}
\author[1]{Alexander E. Beasley}
\author[1,3]{Richard Mayne}

\address[1]{Unconventional Computing Laboratory, UWE, Bristol, UK}
\address[*]{Corresponding author: andrew.adamatzky@uwe.ac.uk}
\address[2]{Microbiology Department, University of Utrecht, Utrecht, The Netherlands}
\address[3]{Department of Applied Sciences,
University of the West of England,
 UWE, Bristol, UK}

\begin{abstract}
\noindent
A fungal colony maintains its integrity via flow of cytoplasm along mycelium network. This flow, together with possible coordination of mycelium tips propagation, is controlled by calcium waves and associated waves of electrical potential changes. We propose that these excitation waves can be employed to implement a computation in the mycelium networks. We use FitzHugh-Nagumo model to imitate propagation of excitation in  a single colony of \emph{Aspergillus niger}. Boolean values are encoded by spikes of extracellular potential. We represent binary inputs by electrical impulses on a pair of selected electrodes and we record responses of the colony from sixteen electrodes. We derive sets of two-inputs-on-output logical gates implementable the fungal colony and analyse distributions of the gates. 
\end{abstract}

\begin{keyword}
  mycelium network, Boolean gates, unconventional computing
\end{keyword}

\end{frontmatter}

\section{Introduction}

A vibrant field of unconventional computing aims to employ space-time dynamics of physical, chemical and biological media to design novel computational techniques, architectures and working prototypes of embedded computing substrates and devices. Interaction-based computing devices, is one of the most diverse and promising families of the unconventional computing structures. They are based on interactions of fluid streams, signals propagating along conductors or excitation wave-fronts, see e.g. \cite{steinbock1996chemical, sielewiesiuk2001logical,adamatzky2004experimental,adamatzky2004experimental, yokoi2004excitable, DBLP:journals/ijuc/Vazquez-OteroFDD14, DBLP:journals/ijuc/IgarashiG11, gorecki2006information, gorecki2009information, stovold2012simulating, gentili2012belousov, takigawa2011dendritic, stovold2012simulating,  gruenert2015understanding}. Typically, logical gates and their cascade implemented in an excitable medium are `handcrafted' to address exact timing and type of interactions between colliding wave-fronts~\cite{steinbock1996chemical, sielewiesiuk2001logical, adamatzky2004collision, adamatzky2007binary, toth2010simple, adamatzky2011towards, de2009implementation, sun2013multi, zhang2012towards, suncrossover, digitalcomparator, adamatzky2011polymorphic, stevens2012time}. The artificial design of logical circuits might be suitable when chemical media or functional materials are used. However, the approach might be not feasible when embedding computation in living systems, where the architecture of conductive pathways may be difficult to alter or control.  In such situations an opportunistic approach to outsourcing computation can be adopted. The system is perturbed via two or more input loci and its dynamics if recorded at one or more output loci. A wave-front appearing at one of the output loci is interpreted as logical truth or `1'. Thus the system with relatively unknown structure implements a mapping $\{0, 1\}^n \rightarrow \{0, 1\}^m$, where $n$ is a number of input loci and $m$ is a number of output loci, $n, m>0$~\cite{adamatzky2019computing,adamatzky2019plant}. The approach belong to same family of computation outsourcing techniques as  \emph{in materio} computing~\cite{miller2002evolution,miller2014evolution,stepney2019co,miller2018materio,miller2019alchemy} and reservoir computing~\cite{verstraeten2007experimental,lukovsevivcius2009reservoir,dale2017reservoir,konkoli2018reservoir,dale2019substrate}.

\begin{figure}[!tbp]
    \centering
    \includegraphics[width=0.4\textwidth]{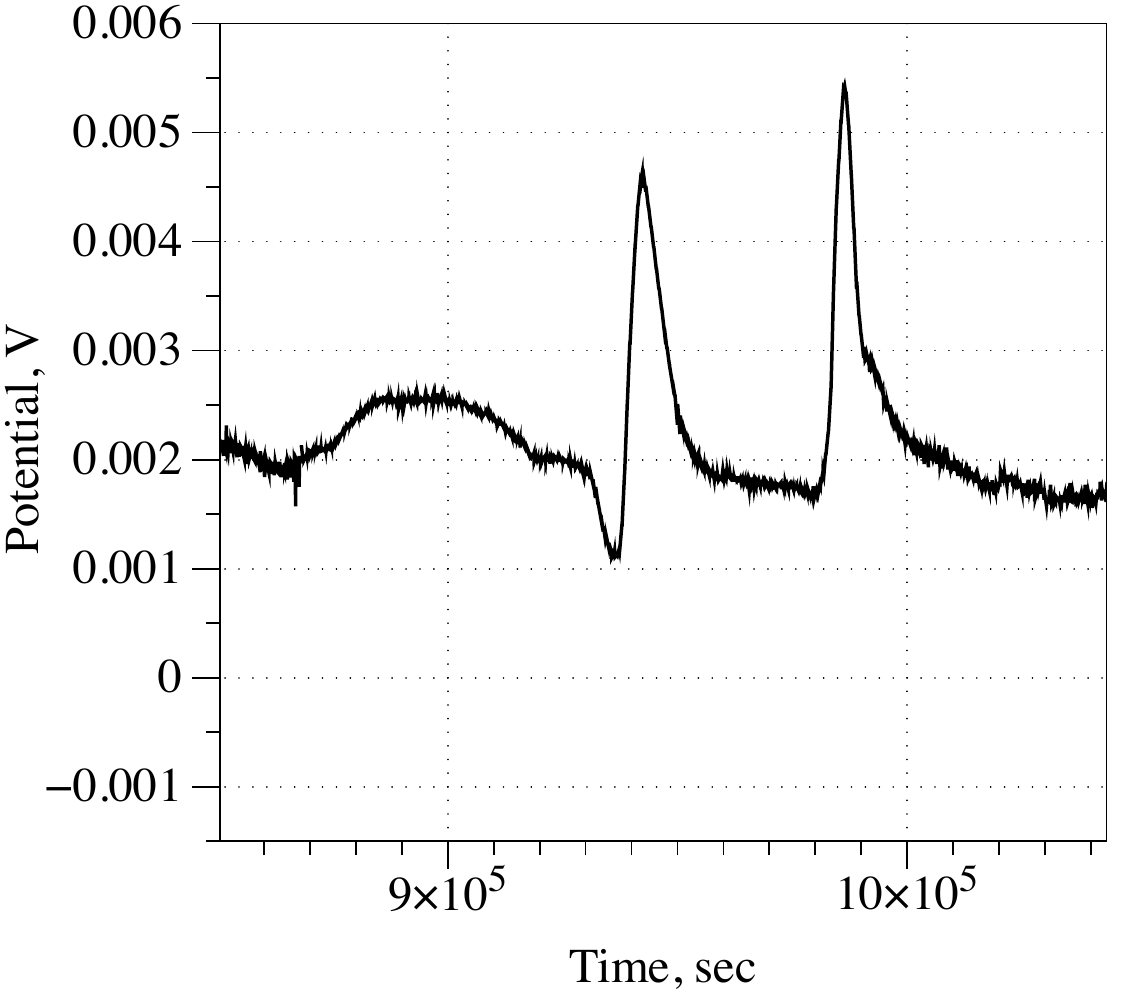}
    \caption{Exemplar spikes of extracellular electrical potential propagating in fungal mycelium.}
    \label{fig:expemplarSpikes}
\end{figure}

Fungal colonies are characterised by rich typology  of mycelium networks~\cite{hitchcock1996image,giovannetti2004patterns,fricker2007network,fricker2017mycelium,islam2017morphology} in some cases affine to fractal structures~\cite{obert1990microbial,patankar1993fractal,bolton1993characterization,mihail1995fractal,boddy1999fractal,papagianni2006quantification}. Rich morphological features might imply rich computational abilities and thus worse to analyse from realising Boolean functions point of view. In numerical experiments we study implementation of logical gates via interaction of numerous travelling excitation waves, seen as as action potentials, on an image of a real fungal colony. Action potential-like spikes of electrical potential have been discovered using intra-cellular recording of mycelium of \emph{Neurospora crassa}~\cite{slayman1976action} and further confirmed in intra-cellular recordings of action potential in hypha of \emph{Pleurotus ostreatus} and \emph{Armillaria bulbosa}~\cite{olsson1995action} and in extra-cellular recordings of fruit bodies of and substrates colonized by mycelium of \emph{Pleurotus ostreatus}~\cite{adamatzky2018spiking} (Fig.~\ref{fig:expemplarSpikes}). While the exact nature of the travelling spikes remains uncertain we can speculate, by drawing analogies with oscillations of electrical potential of slime mould~\emph{Physarum polycephalum}~\cite{iwamura1949correlations, kamiya1950bioelectric, kishimoto1958rhythmicity, meyer1979studies}, that the spikes in fungi are triggered by calcium waves, reversing of cytoplasmic flow, translocation of nutrients and metabolites. Studies of electrical activity of higher plants can brings us even more clues. Thus, the plants use the electrical spikes for a long-distance communication aimed to coordinate an activity of their bodies~\cite{trebacz2006electrical,fromm2007electrical,zimmermann2013electrical}. The spikes of electrical potential in plants relate to a motor activity~\cite{simons1981role,fromm1991control,sibaoka1991rapid,volkov2010mimosa}, responses to changes in temperature~\cite{minorsky1989temperature}, osmotic environment~\cite{volkov2000green} and mechanical stimulation~\cite{roblin1985analysis,pickard1973action}. 
The paper is structured as follows. Colony imaging and numerical solutions of FitzHugh-Nagumo equations are introduced in Sect.~\ref{methods}. Section~\ref{results} studies a role of excitability on the coverage of the network by travelling waves of excitation and exemplifies distributions of Boolean computable on the given mycelium network. We discuss operation characteristics of the mycelium computer in Sect.~\ref{discussion}.

\section{Methods}
\label{methods}

\subsection{Colony imaging}

\emph{Aspergillus niger} strain AR9\#2~\cite{vinck2011heterogenic}, expressing Green Fluorescent Protein (GFP) from the glucoamylase (\emph{glaA}) promoter, was grown at 30\textsuperscript{o}C on minimal medium (MM)~\cite{de2004new} with 25~mM xylose and 1.5\% agarose (MMXA). MMXA cultures were grown for three days, after which conidia were harvested using saline-Tween (0.8\% NaCl and 0.005\% Tween-80). 250~ml liquid cultures were inoculated with $1.25\cdot 10^9$ freshly harvested conidia and grown at 200~rpm and 30\textsuperscript{o}C in 1~L Erlenmeyer flasks in complete medium (CM) (MM containing 0.5\% yeast extract and 0.2\% enzymatically hydrolyzed casein) supplemented with 25~mM xylose (repressing glaA expression). Mycelium was harvested after 16 h and washed twice with PBS. Ten g of biomass (wet weight)  was transferred to MM supplemented with 25~mM maltose (inducing glaA expression). 

Fluorescence of GFP was localised in micro-colonies using a DMI 6000 CS AFC confocal microscope (Leica, Mannheim, Germany). Micro-colonies were fixed overnight at 4\textsuperscript{o}C in 4\% paraformaldehyde in PBS, washed twice with PBS and taken up in 50~ml PBS supplemented with 150~mM glycine to quench autofluorescence. Micro-colonies were then transferred to a glass bottom dish (Cellview\texttrademark, Greiner Bio-One, Frickenhausen, Germany, PS, 35/10 MM) and embedded in 1\% low melting point agarose at 45\textsuperscript{o}C. Micro-colonies were imaged at 20$\times$ magnification (HC PL FLUOTAR L 20 $\times$ 0.40 DRY). GFP was excited by white light laser at 472~nm using 50\% laser intensity (0.1~kW/cm2) and a pixel dwell time of 72~ns. Fluorescent light emission was detected with hybrid detectors in the range of 490–525~nm. Pinhole size was 1 Airy unit. Z-stacks of imaged micro-colonies were made using 100 slices with a slice thickness of 8.35~µm.
3D projections were made with Fiji~\cite{schindelin2012fiji}.

\subsection{Numerical modelling}

\begin{figure}[!tbp]
    \centering
    \subfigure[]{\includegraphics[width=0.49\textwidth]{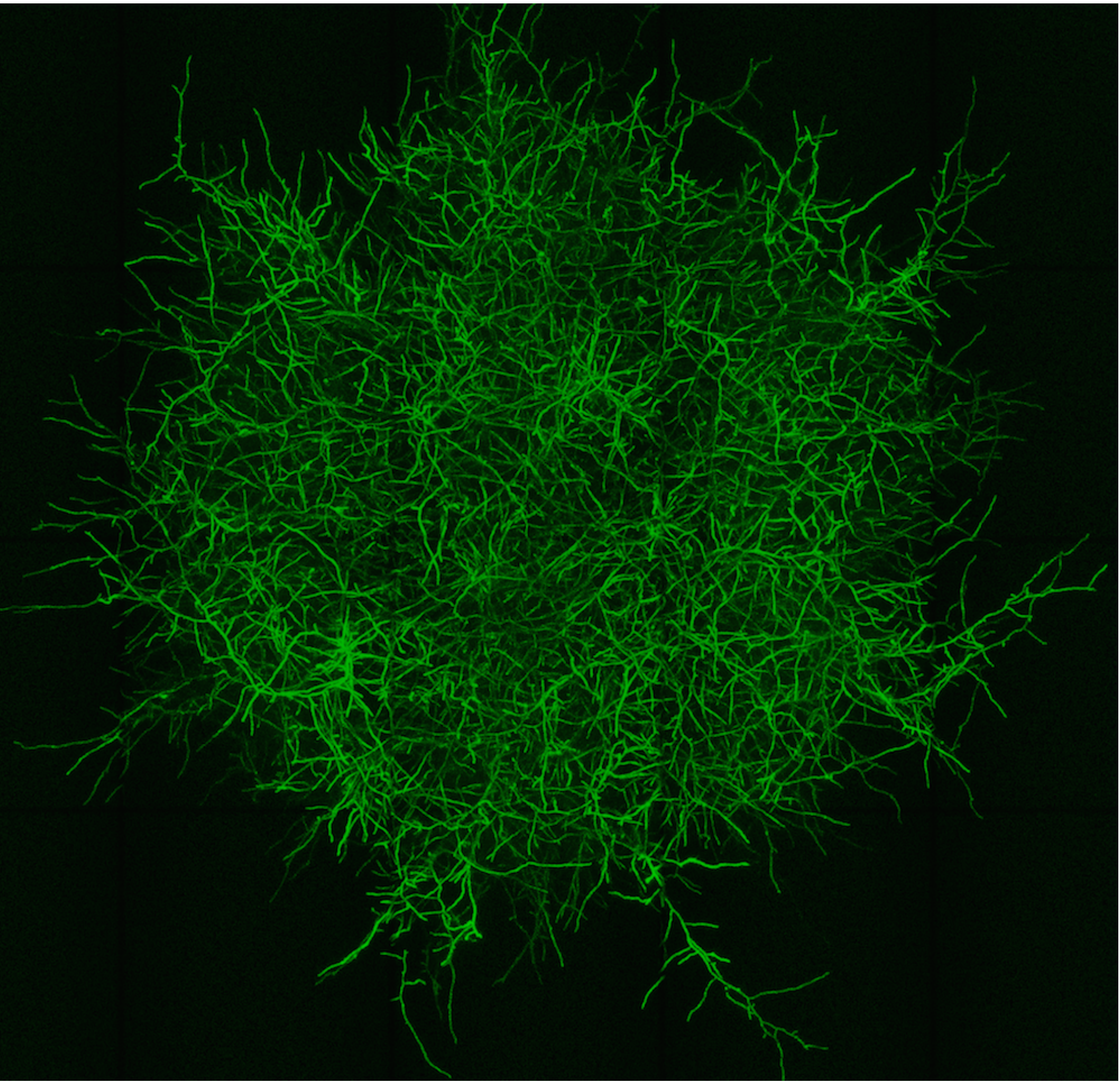}}
    \subfigure[]{\includegraphics[width=0.49\textwidth]{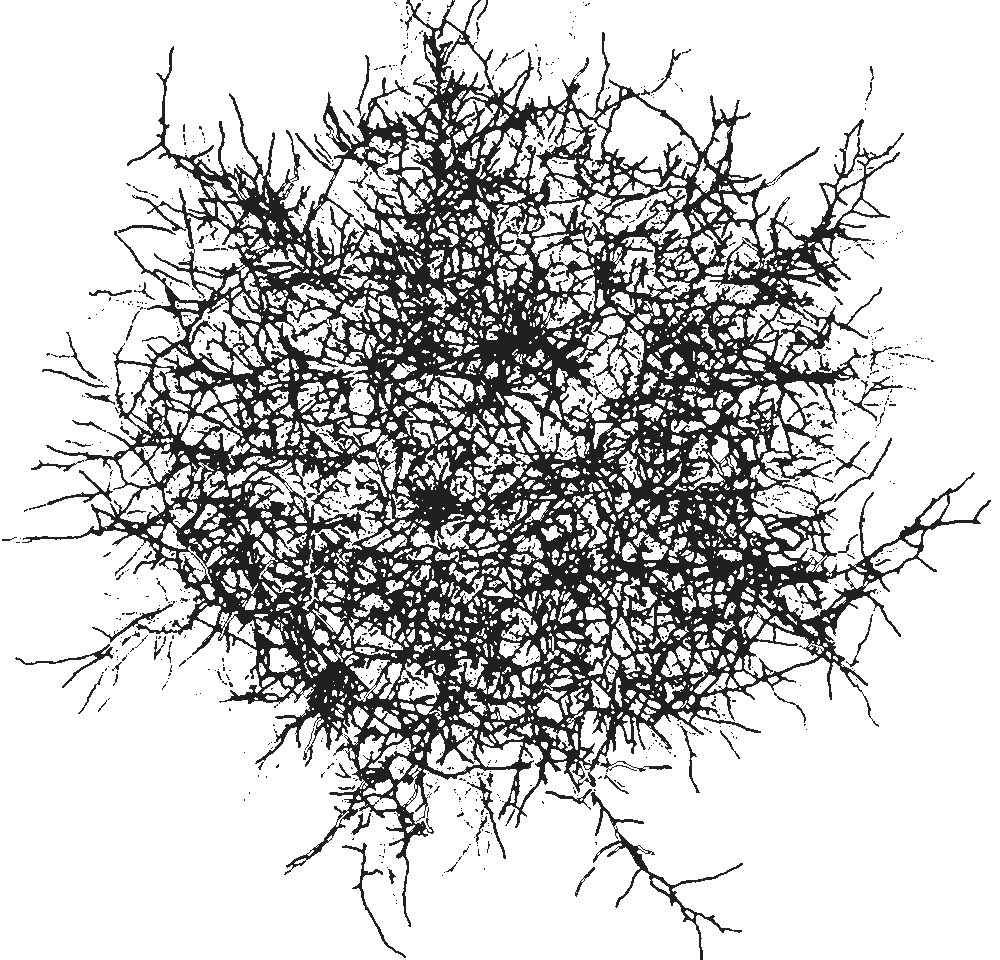}}
     \subfigure[]{\includegraphics[width=0.49\textwidth]{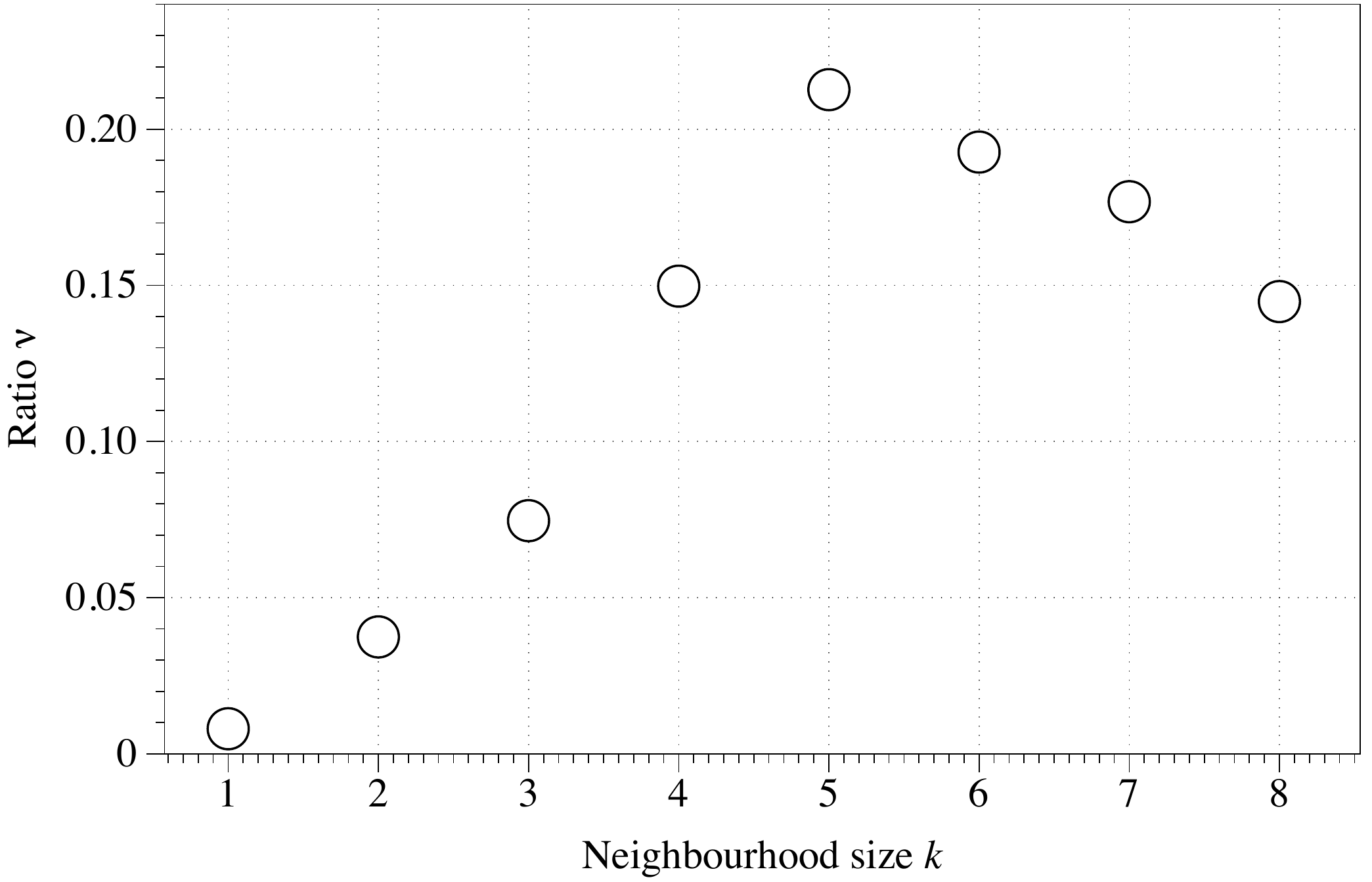}}
      \subfigure[]{\includegraphics[width=0.49\textwidth]{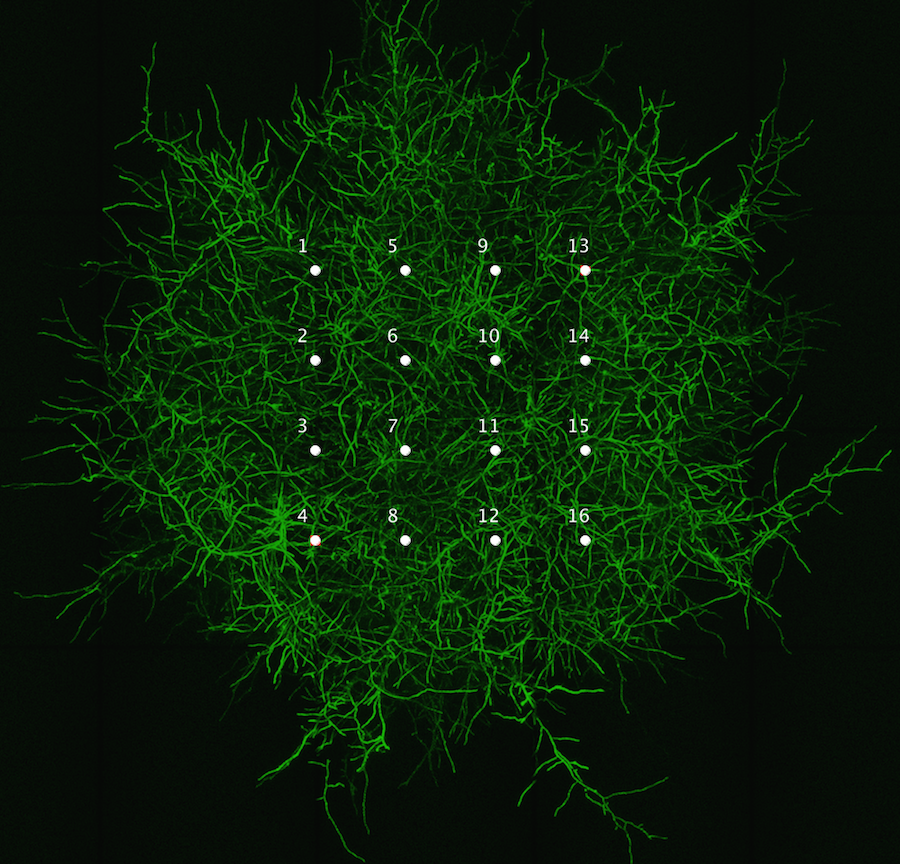}}
    \caption{Image of the fungal colony, $1000 \times 960$ pixels used as a template conductive for FHN. 
    (a)~Original image, mycelium is seen as green pixels. (b)~Conductive matrix $C$, conductive pixels are black.
    (c)~Distribution of neigbourhood sizes.
    (d)~Configuration of electrodes.}
    \label{fig:mycelium}
\end{figure}

We used still image of the colony as a conductive template. The image of the fungal colony  (Fig.~\ref{fig:mycelium}a) was projected onto a $1000 \times 960$ nodes grid. 
The original image $M=(m_{ij})_{1 \leq j \leq n_i, 1 \leq j \leq n_j}$, $m_{ij} \in \{ r_{ij}, g_{ij}, b_{ij} \}$, where $n_i=1000$ and $n_j=960$, and $1 \leq r, g, b \leq 255$ (Fig.~\ref{fig:mycelium}a), was converted to a conductive matrix $C=(m_{ij})_{1 \leq i,j \leq n}$ (Fig.~\ref{fig:mycelium}b) derived from the image as follows: $m_{ij}=1$  if $r_{ij}<20$, $(g_{ij}>40)$ and $b_{ij}<20$; a dilution operation was applied to $C$. 

FitzHugh-Nagumo (FHN) equations~\cite{fitzhugh1961impulses,nagumo1962active,pertsov1993spiral} {is} a qualitative approximation of the Hodgkin-Huxley model~\cite{beeler1977reconstruction} of electrical activity of living cells:
\begin{eqnarray}
\frac{\partial v}{\partial t} & = & c_1 u (u-a) (1-u) - c_2 u v + I + D_u \nabla^2 \\
\frac{\partial v}{\partial t} & = & b (u - v),
\end{eqnarray}
where $u$ is a value of a trans-membrane potential, $v$ a variable accountable for a total slow ionic current, or a recovery variable responsible for a slow negative feedback, $I$ {is} a value of an external stimulation current. The current through intra-cellular spaces is approximated by
$D_u \nabla^2$, where $D_u$ is a conductance. Detailed explanations of the `mechanics' of the model are provided in~\cite{rogers1994collocation}, here we shortly repeat some insights. The term $D_u \nabla^2 u$ governs a passive spread of the current. The terms $c_2 u (u-a) (1-u)$ and $b (u - v)$ describe the ionic currents. The term $u (u-a) (1-u)$ has two stable fixed points $u=0$ and $u=1$ and one unstable point $u=a$, where $a$ is a threshold of an excitation.

We integrated the system using the Euler method with the five-node Laplace operator, a time step $\Delta t=0.015$ and a grid point spacing $\Delta x = 2$, while other parameters were $D_u=1$, $a=0.13$, $b=0.013$, $c_1=0.26$. We controlled excitability of the medium by varying $c_2$ from 0.05 (fully excitable) to 0.015 (non excitable). Boundaries are considered to be impermeable: $\partial u/\partial \mathbf{n}=0$, where $\mathbf{n}$ is a vector normal to the boundary. 

The waves of excitation propagated on conductive nodes of the grid of $C$, in addition to the parameter $c_2$, excitability of each conductive node was dependent on a number $k$ of its immediate conductive neighbours. Distribution of neighbourhood sizes are shown in Fig.~\ref{fig:mycelium}c.

To show dynamics of excitation in the network we simulated electrodes by calculating a potential $p^t_x$ at an electrode location $x$ as $p_x = \sum_{y: |x-y|<2} (u_x - v_x)$. Configuration of electrodes $1, \cdots, 16$ is shown in Fig.~\ref{fig:mycelium}d. The numerical integration code written in Processing  was inspired by previous methods of numerical integration of FHN and our own computational studies of the impulse propagation in biological networks~\cite{hammer2009, pertsov1993spiral,rogers1994collocation,adamatzky2019interplay,adamatzky2019plant}. Time-lapse snapshots provided in the paper were recorded at every 100\textsuperscript{th} time step, and we display sites with $u >0.04$; videos and figures were produced by saving a frame of the simulation every 100\textsuperscript{th} step of the numerical integration and assembling the saved frames into the video with a play rate of 30 fps. Videos are available at \cite{FTHMyceliumZenodo}.

\section{Results}
\label{results}

While implementing numerical experiments were selected a range the network excitability (Subsect.~\ref{excitability}) and then realised sets of logical gates for excitability values selected (Subsect.~\ref{gates}).

\subsection{Effect of excitability on overall activity}
\label{excitability}

\begin{figure}[!tbp]
    \centering
    \subfigure[$c_2=0.094$]{\includegraphics[scale=0.5]{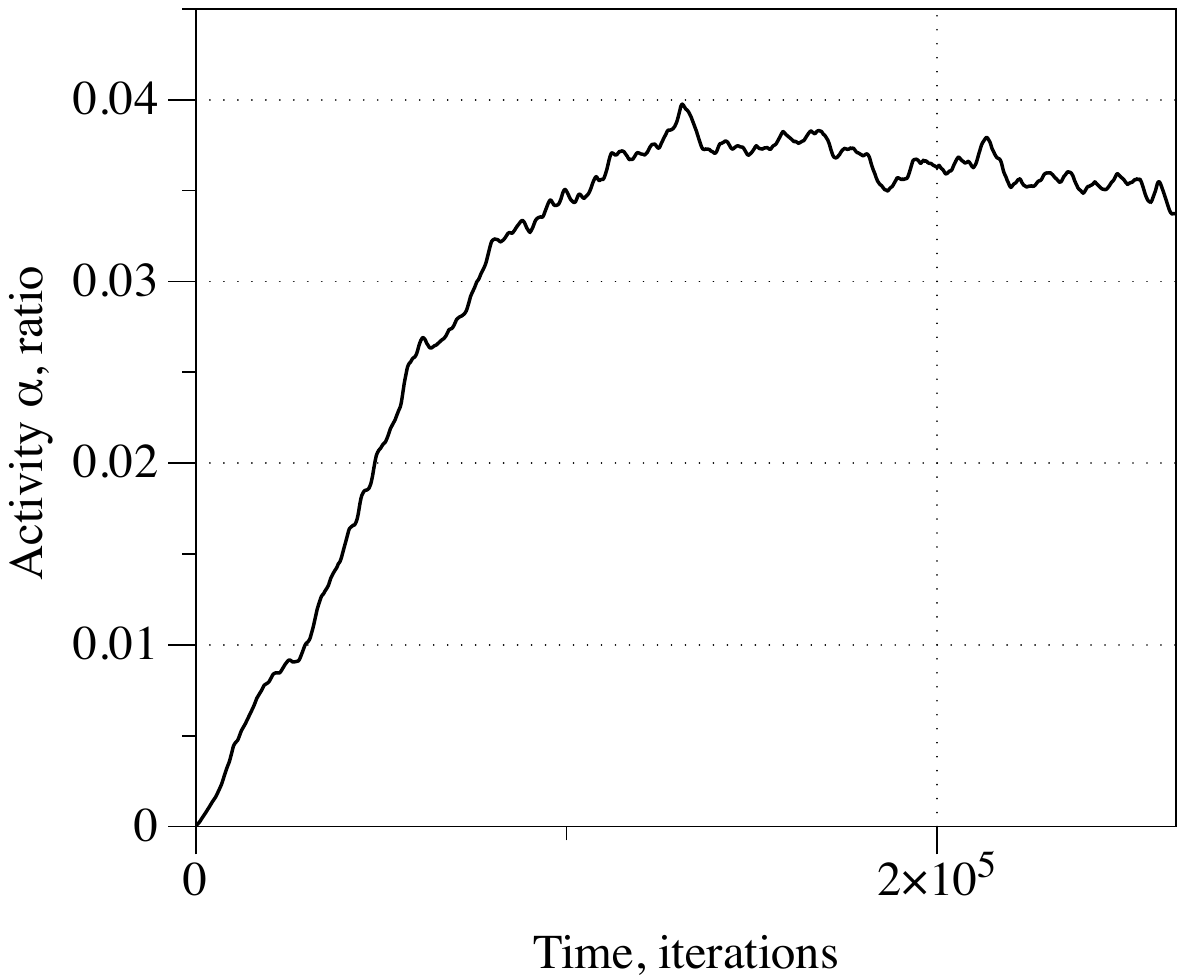}}
    \subfigure[$c_2=0.097$]{\includegraphics[scale=0.5]{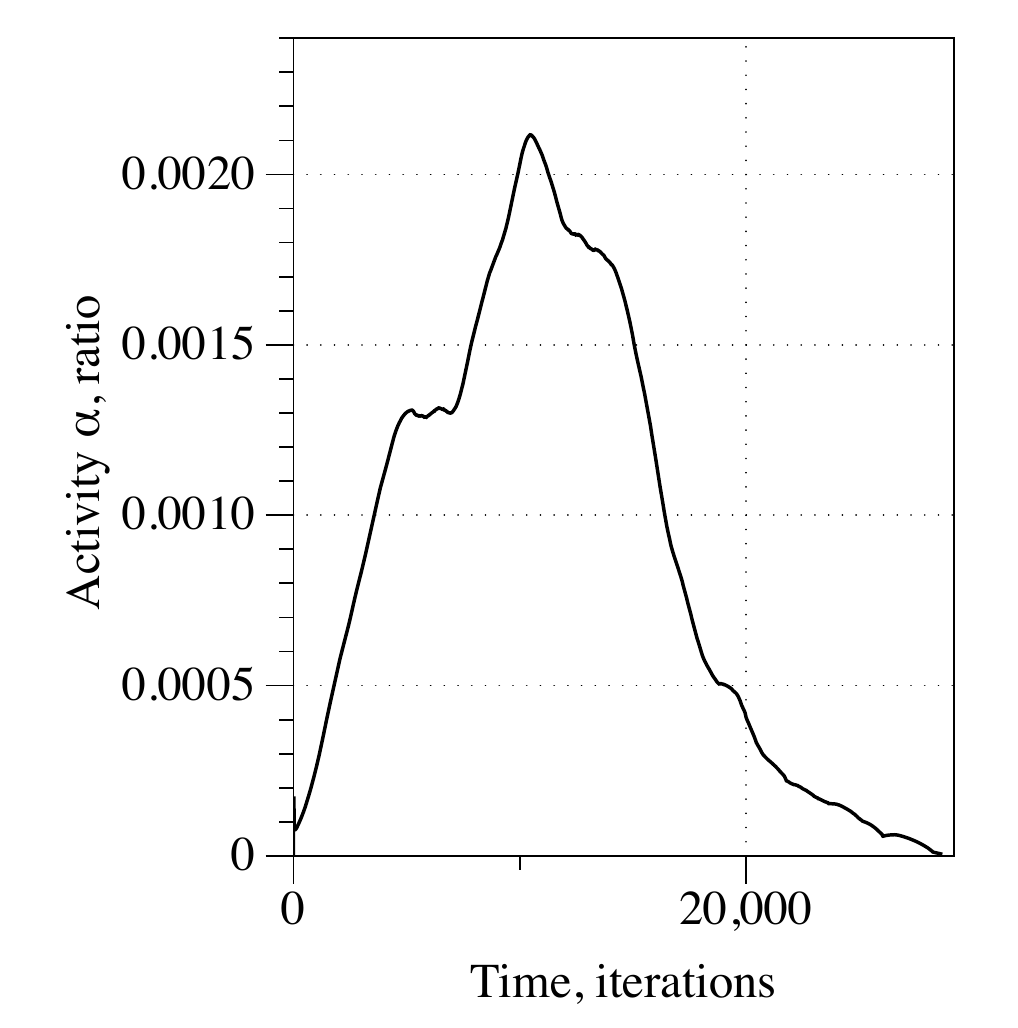}}
    \subfigure[$c_2=0.095$]{\includegraphics[scale=0.5]{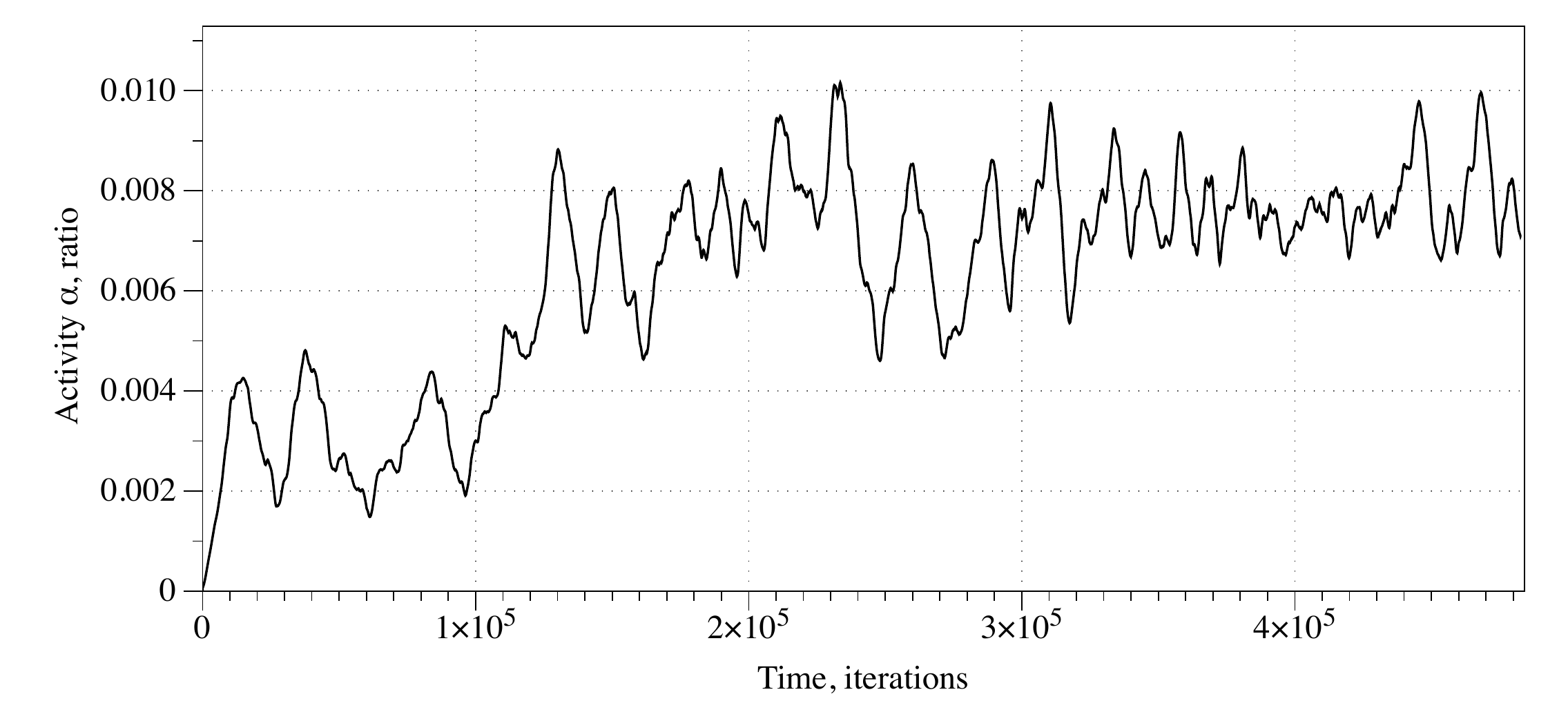}}
\subfigure[$c_2=0.096$]{\includegraphics[scale=0.5]{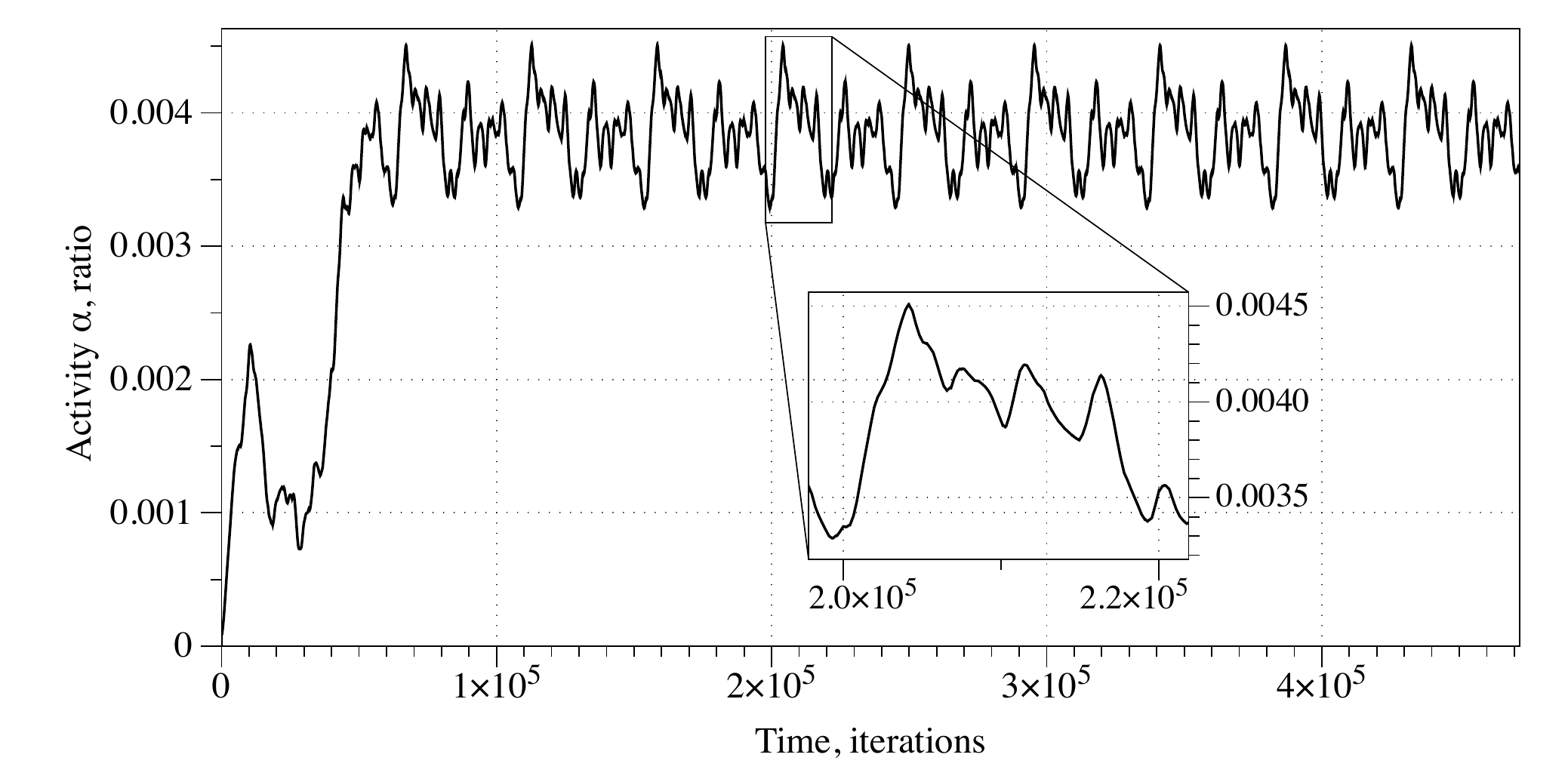}}
    \caption{Dynamics of the activity $\alpha$ for various values of excitability $c_2$, the values are shown in sub-captions.
    For every iteration $t$ we measured the activity of the network as a number of conductive nodes $x$ with $u^t_x>0.1$.}
    \label{fig:activityExamples}
\end{figure}

\begin{figure}[!tbp]
    \centering
    \subfigure[$t=200$]{\includegraphics[width=0.32\textwidth]{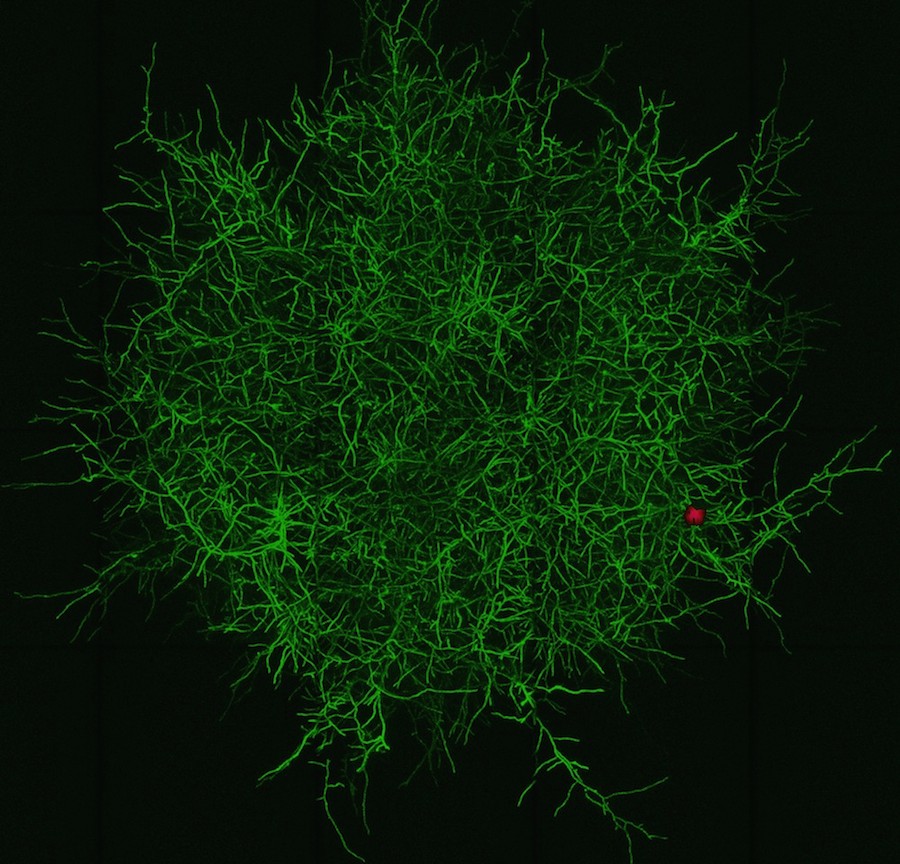}}
    \subfigure[$t=1900$]{\includegraphics[width=0.32\textwidth]{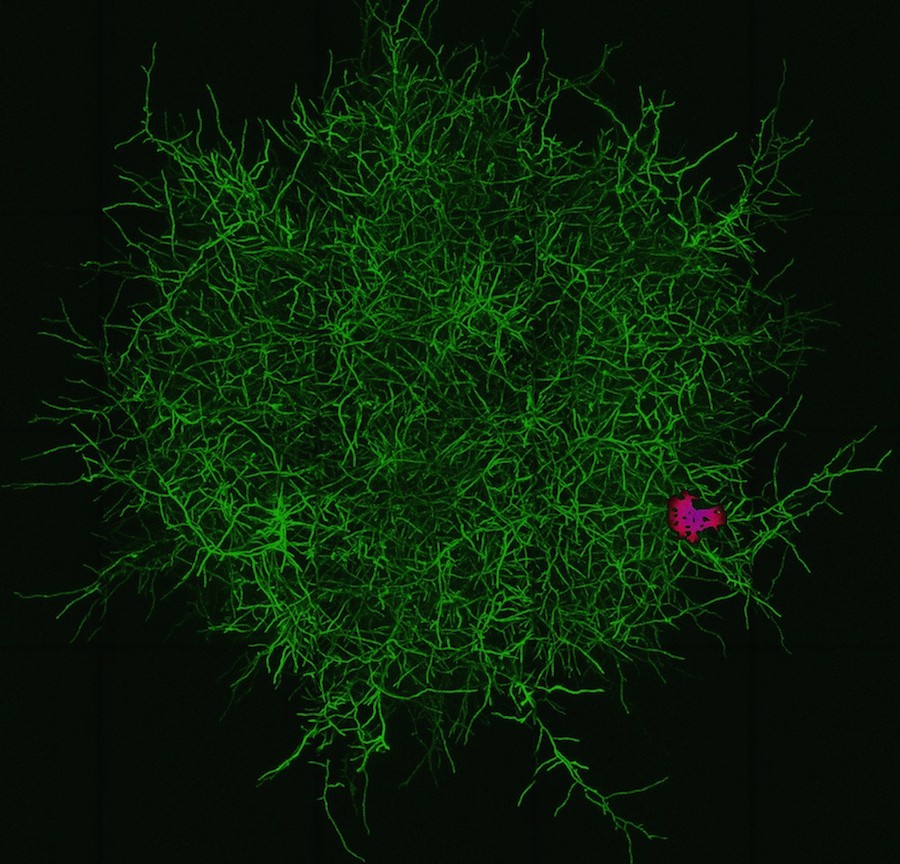}}
    \subfigure[$t=40500$]{\includegraphics[width=0.32\textwidth]{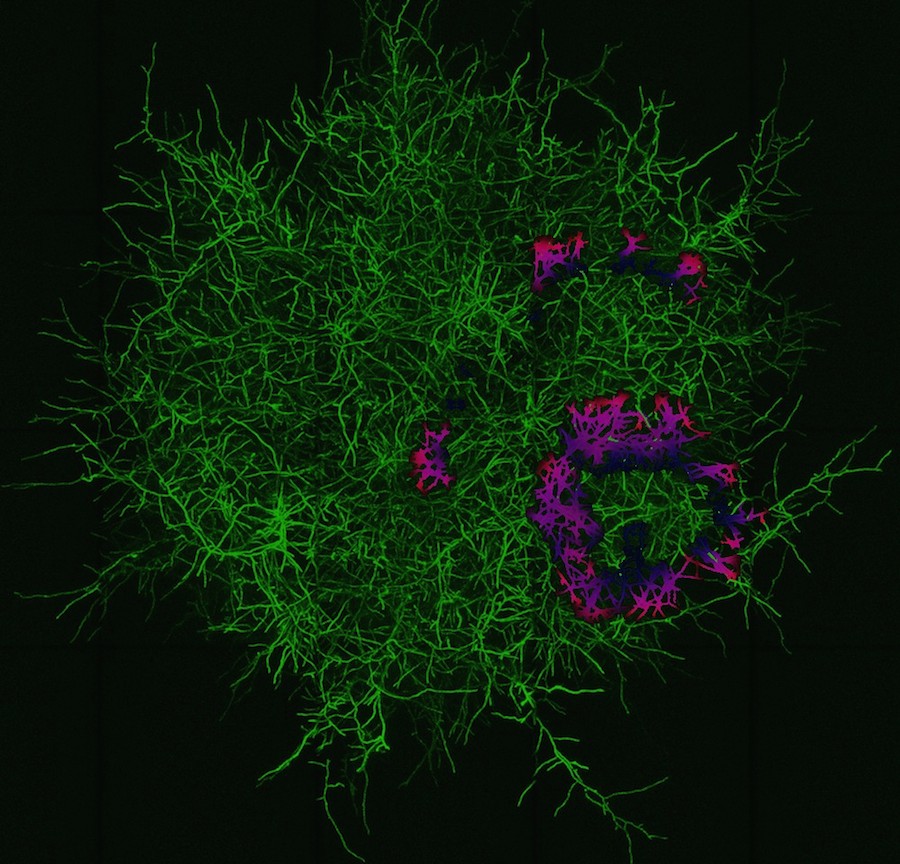}}
    \subfigure[$t=70400$]{\includegraphics[width=0.32\textwidth]{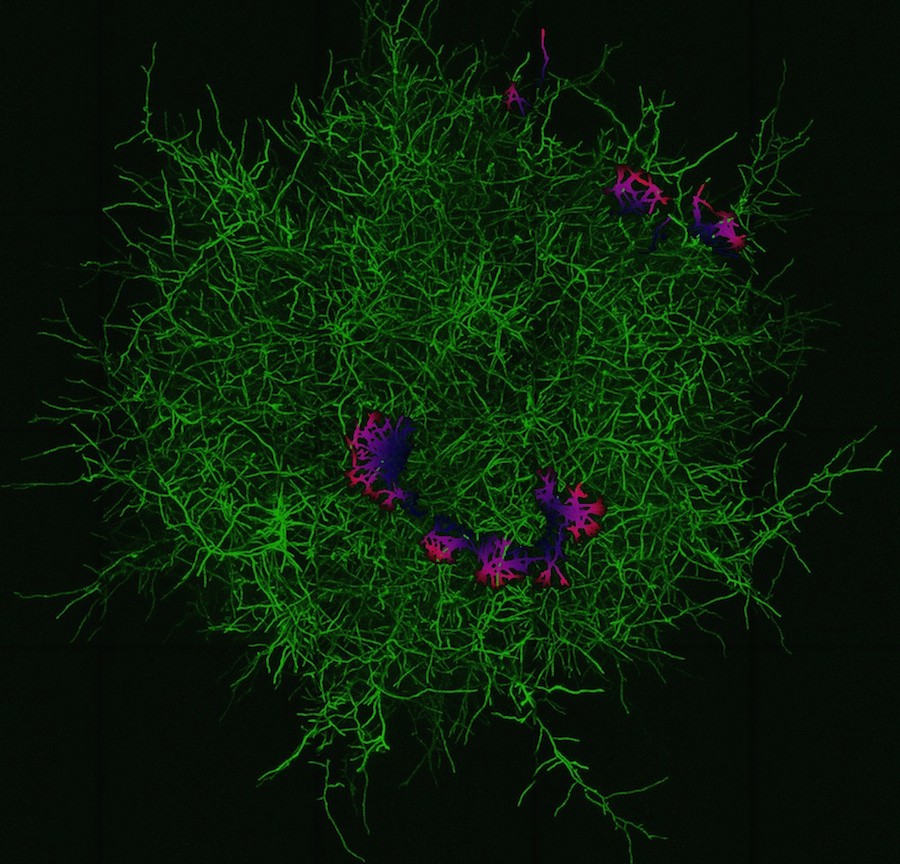}}
    \subfigure[$t=104800$]{\includegraphics[width=0.32\textwidth]{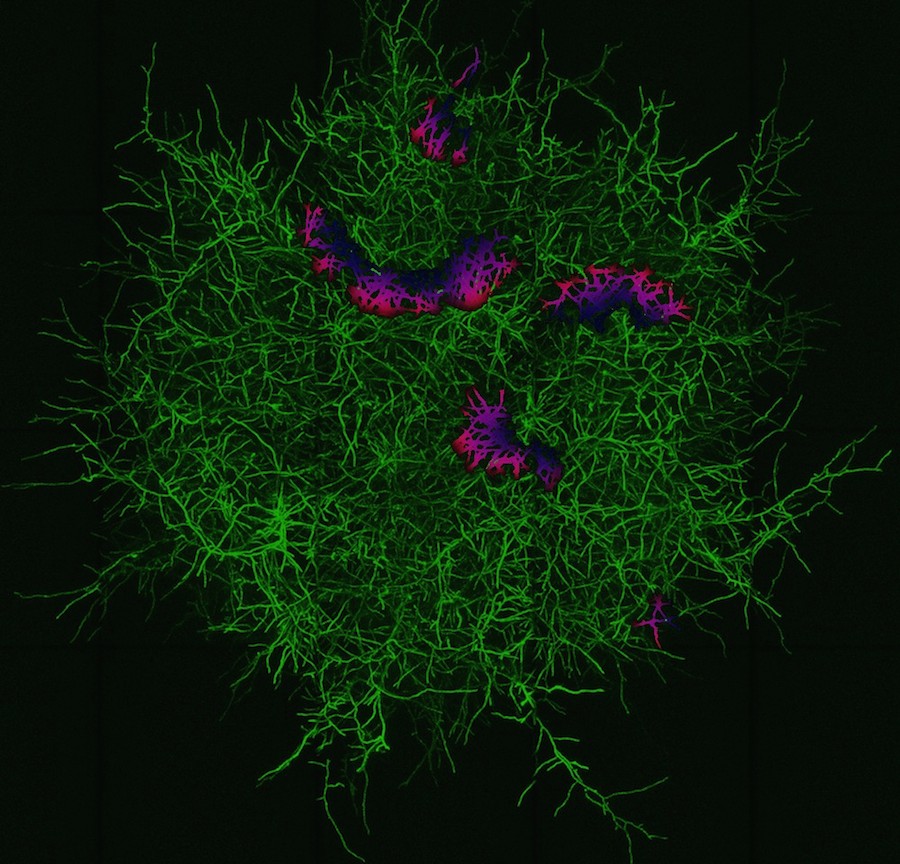}}
    \subfigure[$t=115850$]{\includegraphics[width=0.32\textwidth]{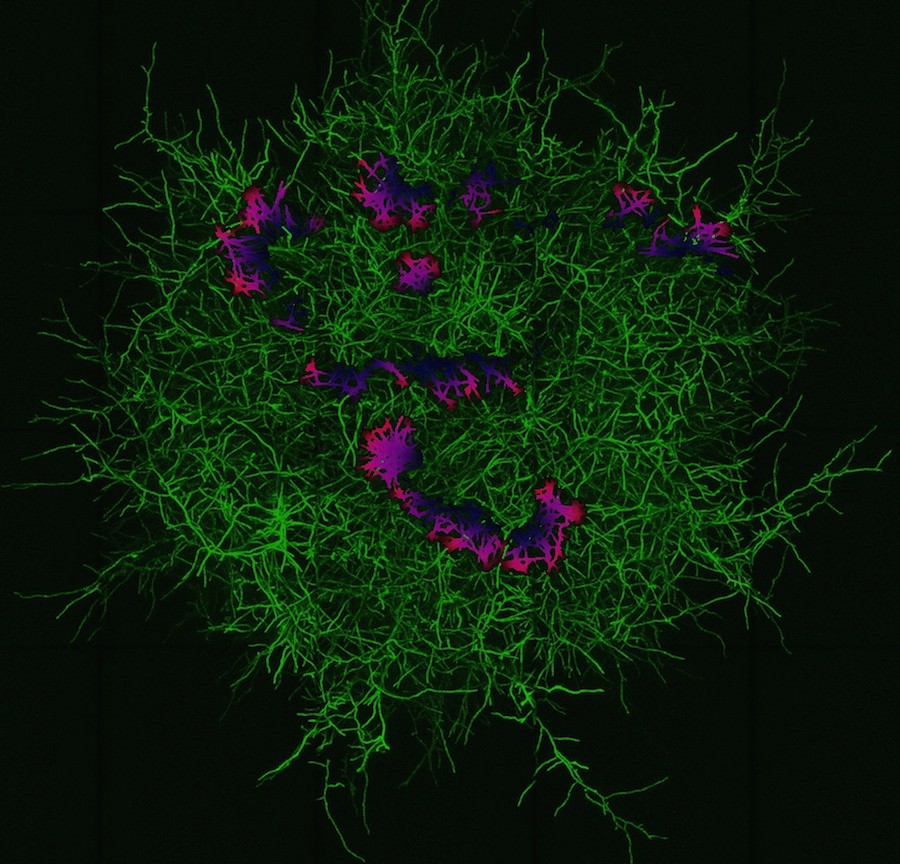}}
    \caption{Snapshots of excitation dynamics for $c_2=0.095$.}
    \label{fig:examples_c2_0095}
\end{figure}

\begin{figure}[!tbp]
    \centering
    \subfigure[$t=1900$]{\includegraphics[width=0.32\textwidth]{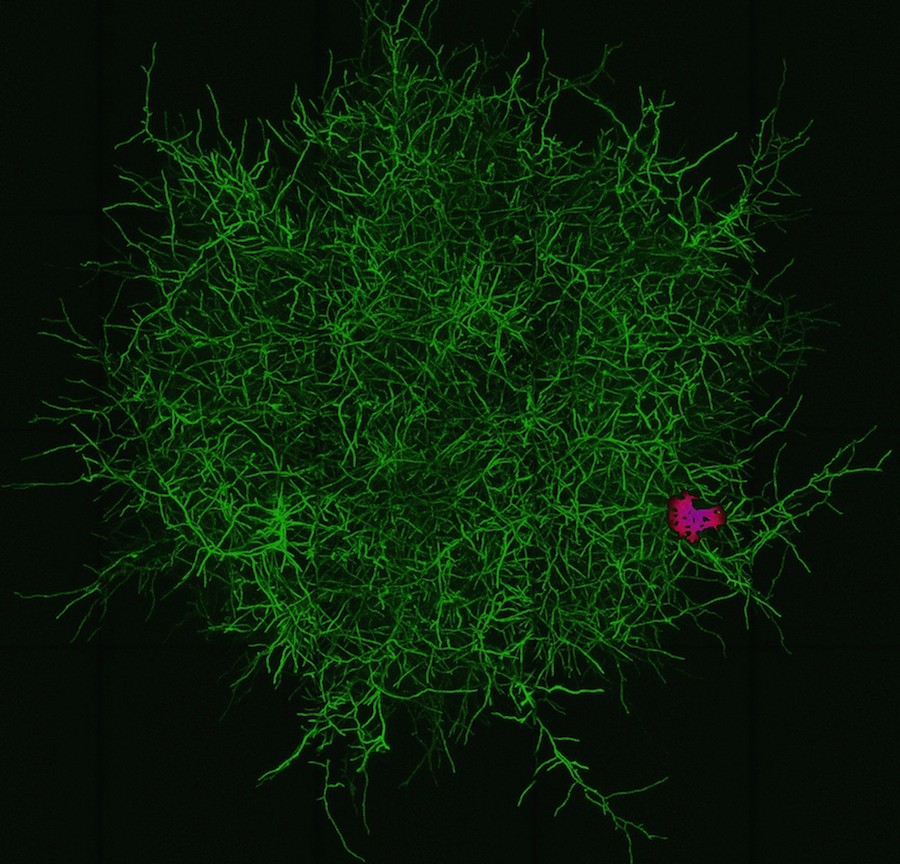}}
    \subfigure[$t=40500$]{\includegraphics[width=0.32\textwidth]{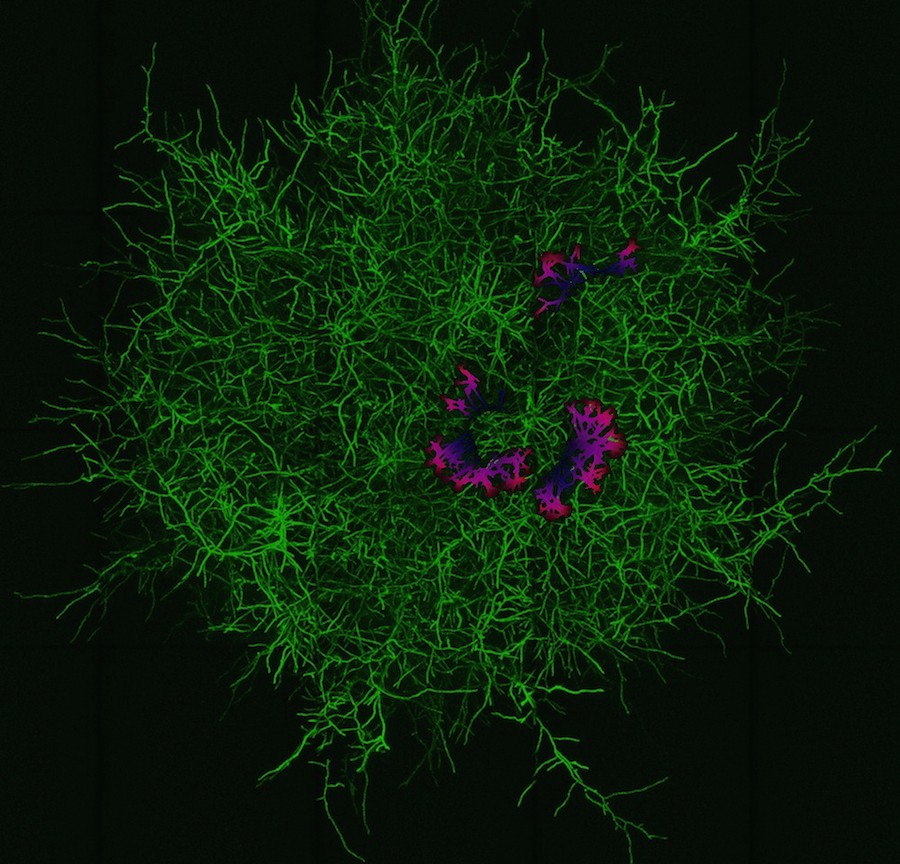}}
    \subfigure[$t=70400$]{\includegraphics[width=0.32\textwidth]{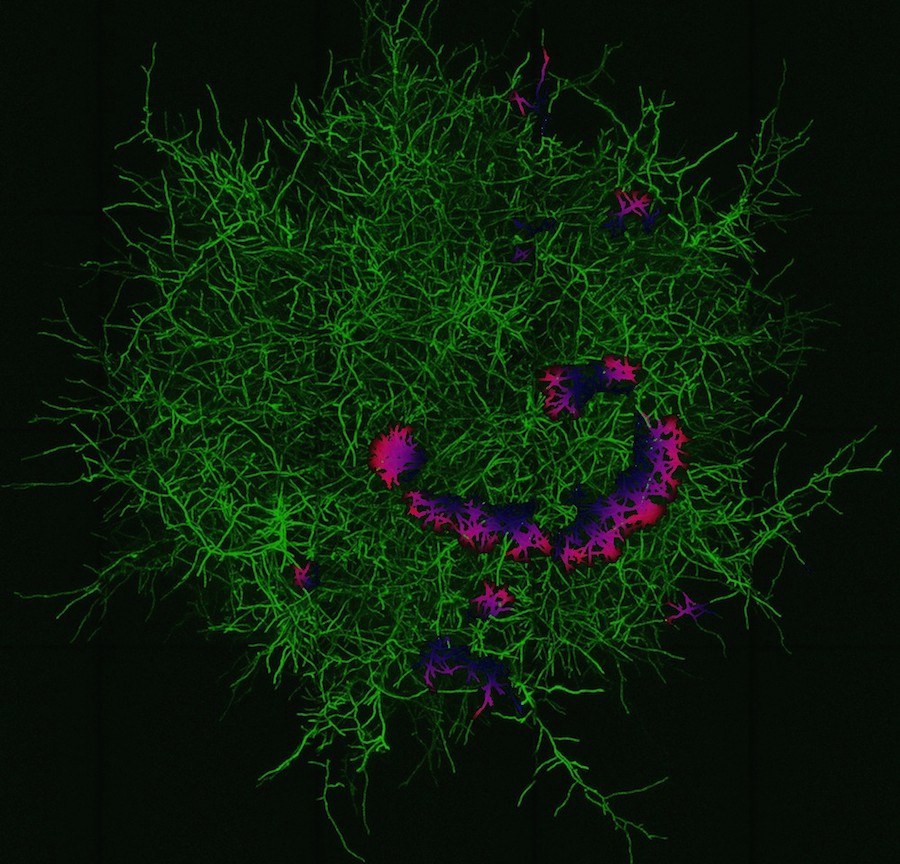}}
    \subfigure[$t=104800$]{\includegraphics[width=0.32\textwidth]{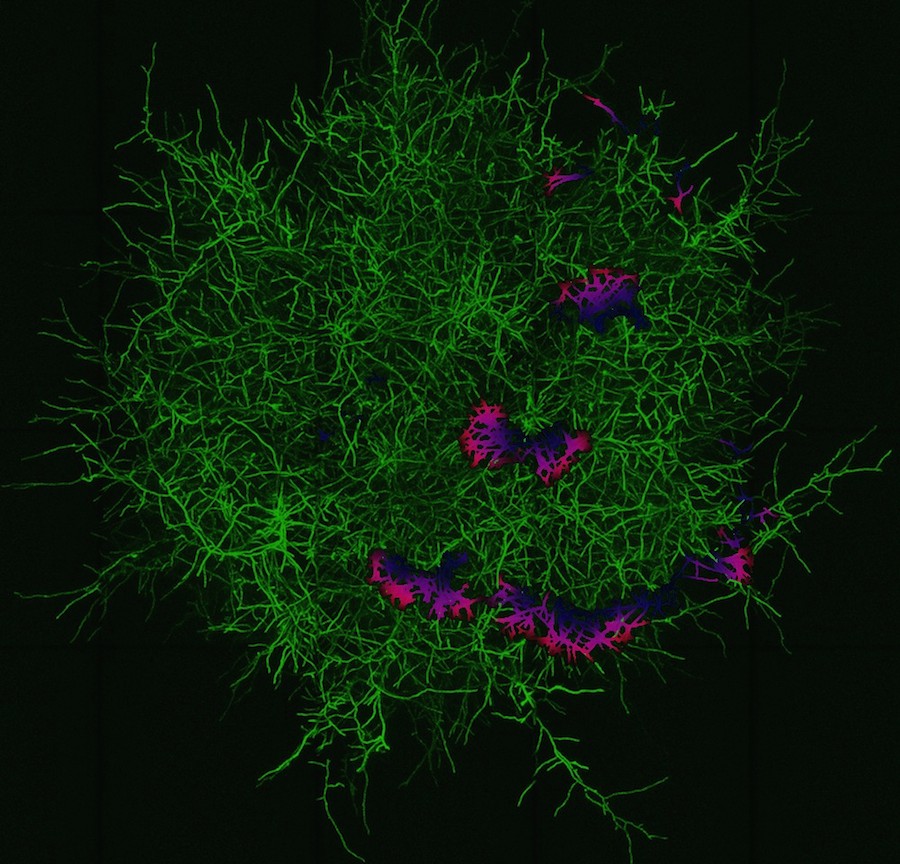}}
    \subfigure[$t=115850$]{\includegraphics[width=0.32\textwidth]{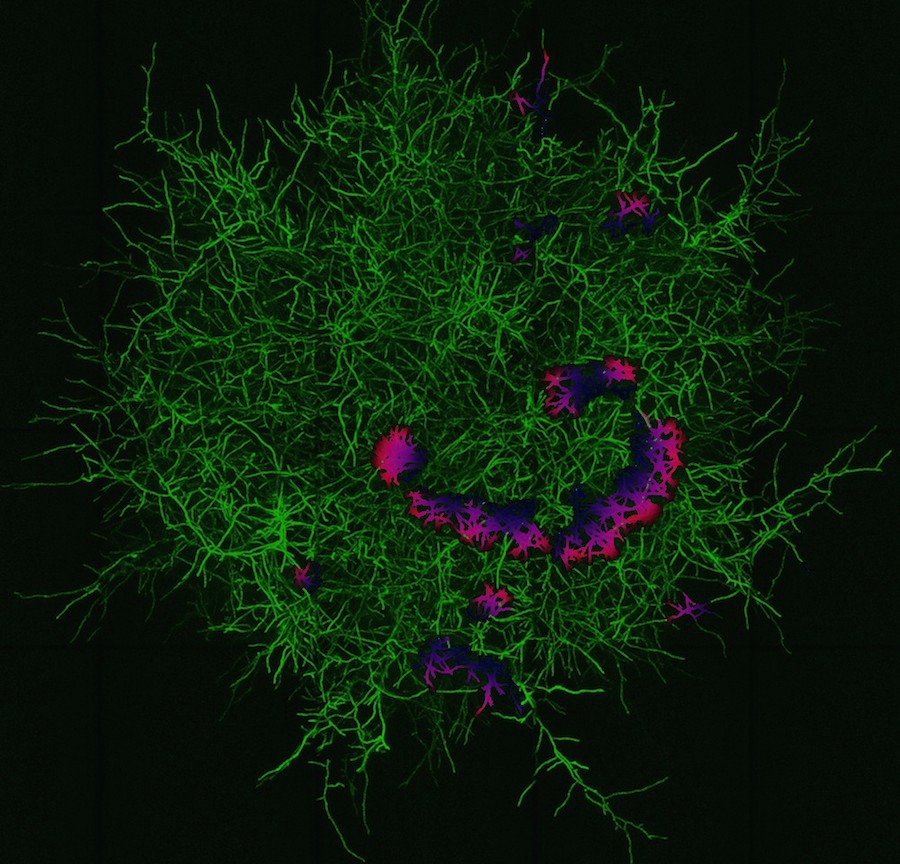}}
     \subfigure[$t=155000$]{\includegraphics[width=0.32\textwidth]{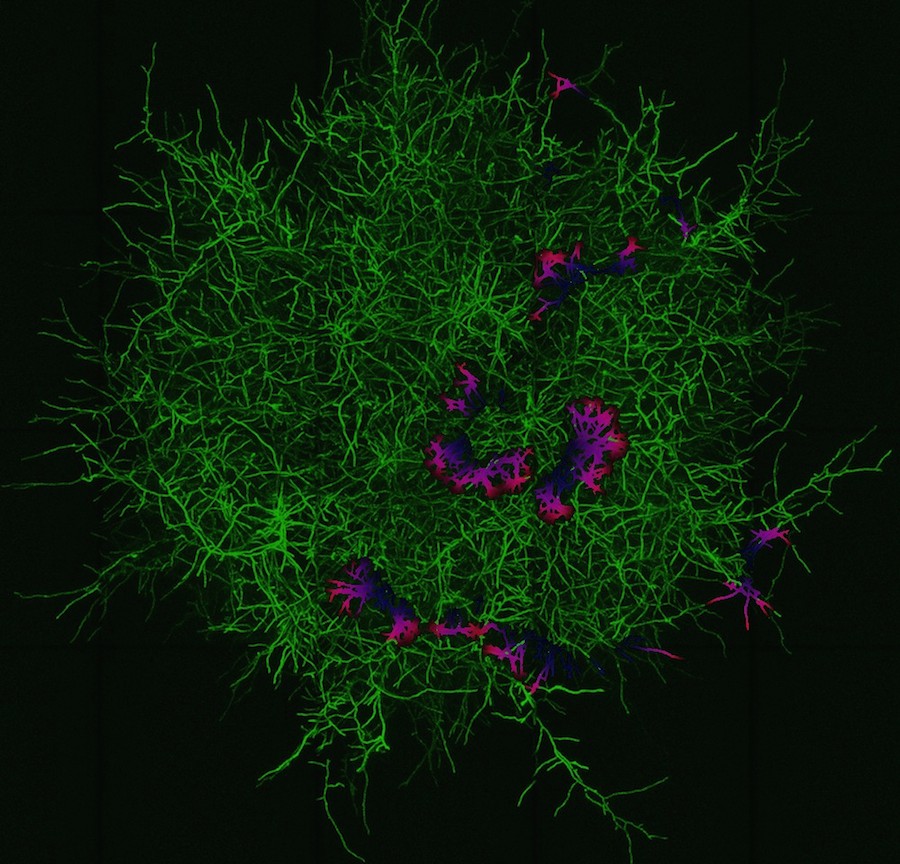}}
    \caption{Snapshots of excitation dynamics for $c_2=0.096$.  Compare (c) and (e): the pattern of excitation returns to the exact point of the cycle.}
    \label{fig:examples_c2_0096}
\end{figure}

\begin{figure}[!tbp]
    \centering
\subfigure[$c_2=0.095$]{\includegraphics[width=0.32\textwidth]{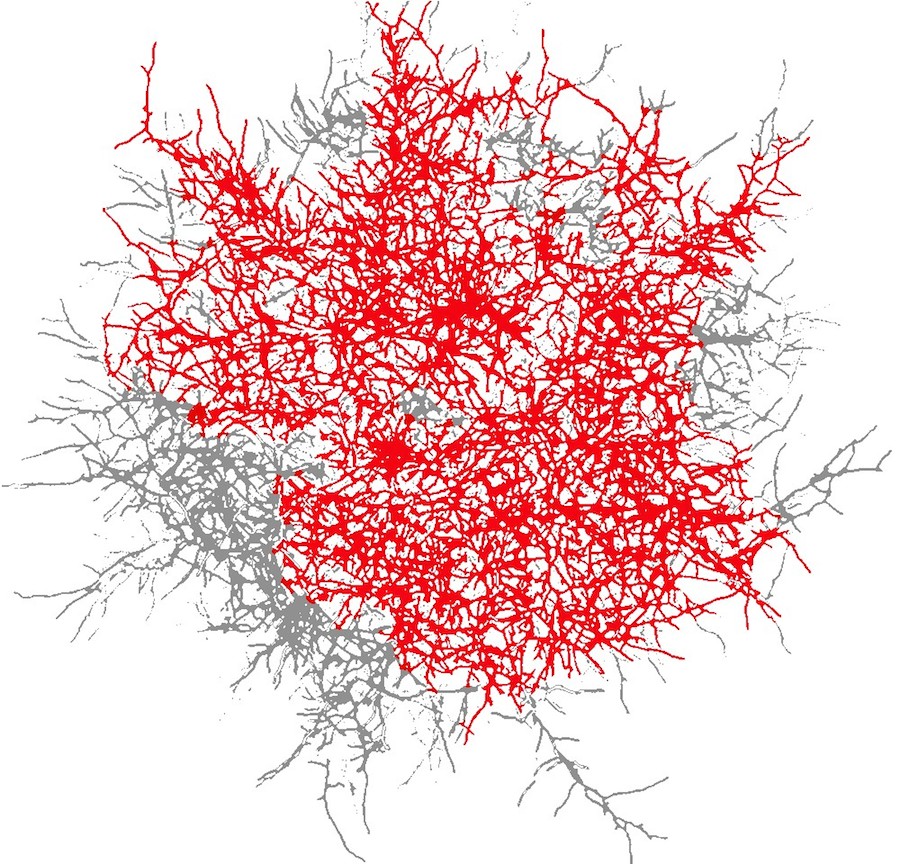}}
\subfigure[$c_2=0.096$]{\includegraphics[width=0.32\textwidth]{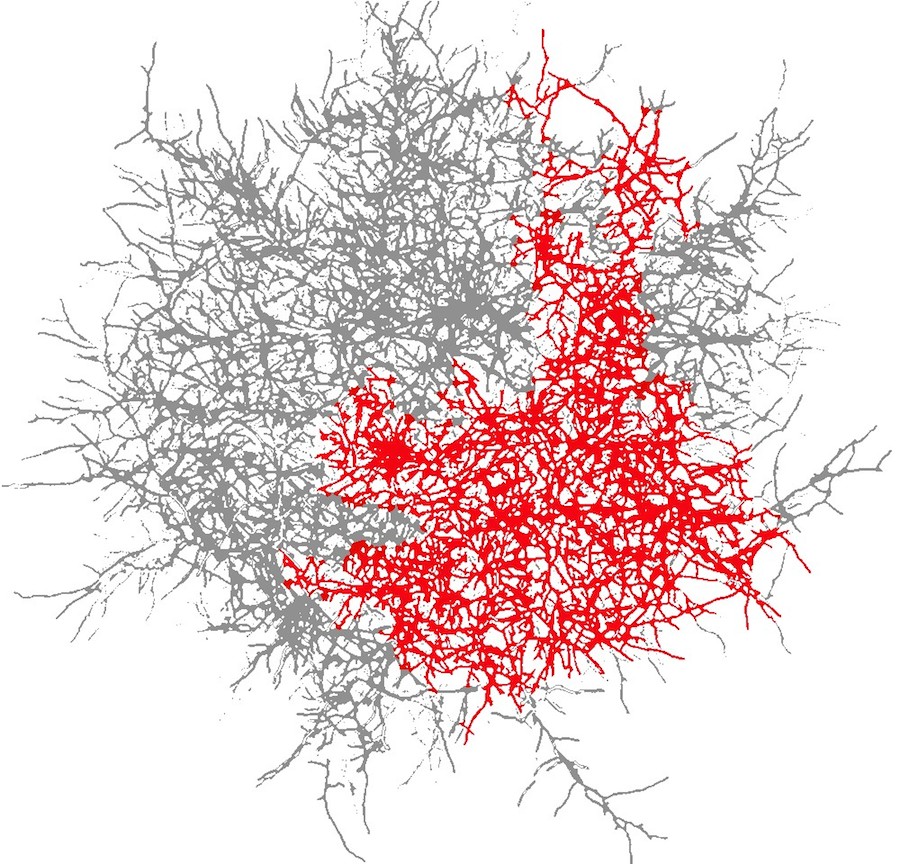}}
\subfigure[$c_2=0.097$]{\includegraphics[width=0.32\textwidth]{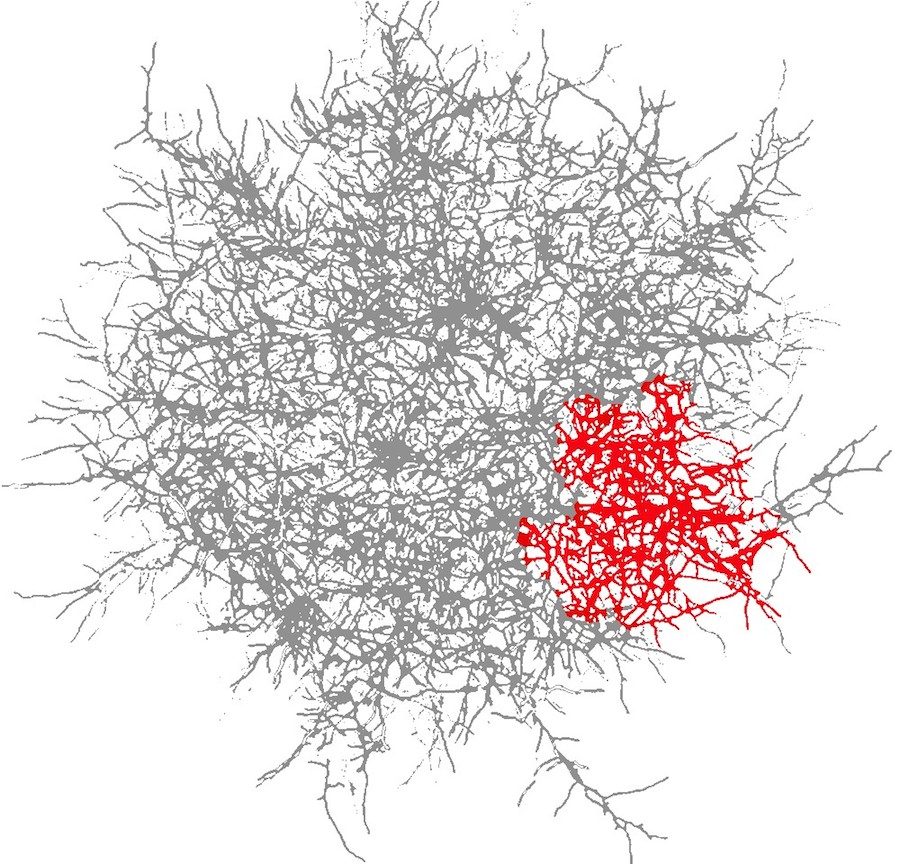}}
    \caption{Coverage of the network for excitability $c_2$ (a)~0.095, (b)~0.096, (c)~0.097. If the a pixel $p$ of the image was excited, $u^t_p>0.1$, it is assumed to be covered and coloured red in the pictures (abc); the pixels which never were excited are coloured gray.}
    \label{fig:coverage}
\end{figure}

For $c_2<0.0945$ any source of excitation  triggers excitation dynamics which occupies all parts of the network accessible, via mycelial strands, from the source. Due to the high level of excitability the network remains in the excitable state (Fig.~\ref{fig:activityExamples}a). For values $c_2$ from 0.094 to 0.00965 we observe propagation of `classical' excitation wave-fronts resembling circular, target and spiral waves in a continuous medium. Examples of wave-fronts propagating in networks with excitability levels $c_2=0.095$ and $c_2=0.096$, excited at the same loci shown in Fig.~\ref{fig:examples_c2_0095}a,  are shown in Figs.\ref{fig:examples_c2_0095} and \ref{fig:examples_c2_0096}. In the network with $c_2$ there are many pathways for propagation of the excitation wave-fronts, therefore, despite being fully deterministic, the network exhibit disordered oscillations of its activity (Fig.~\ref{fig:activityExamples}c). A number of conductive pathways decreases when $c_2$ increases from $0.095$ to $0.096$. Thus many propagating wave-fronts become, relatively, quickly confined to a limited domains of the network, where they continue `circling' indefinitely. A set of regular oscillations of activity becomes evidence after a number of iterations (Fig.~\ref{fig:activityExamples}d). The coverage of the network by excitation wave-fronts reduced with increase of $c_2$ from 0.095 to 0.096 (Fig.~\ref{fig:coverage}ab) and becomes localised when $c_2$ reaches 0.097 (Figs.~\ref{fig:coverage}c and \ref{fig:activityExamples}b). Thus we used networks with $c_2=0.095$ or $0.096$ for implementation of Boolean functions.

\subsection{Distribution of Boolean gates}
\label{gates}

\begin{figure}[!tbp]
    \centering
    \includegraphics[scale=0.6]{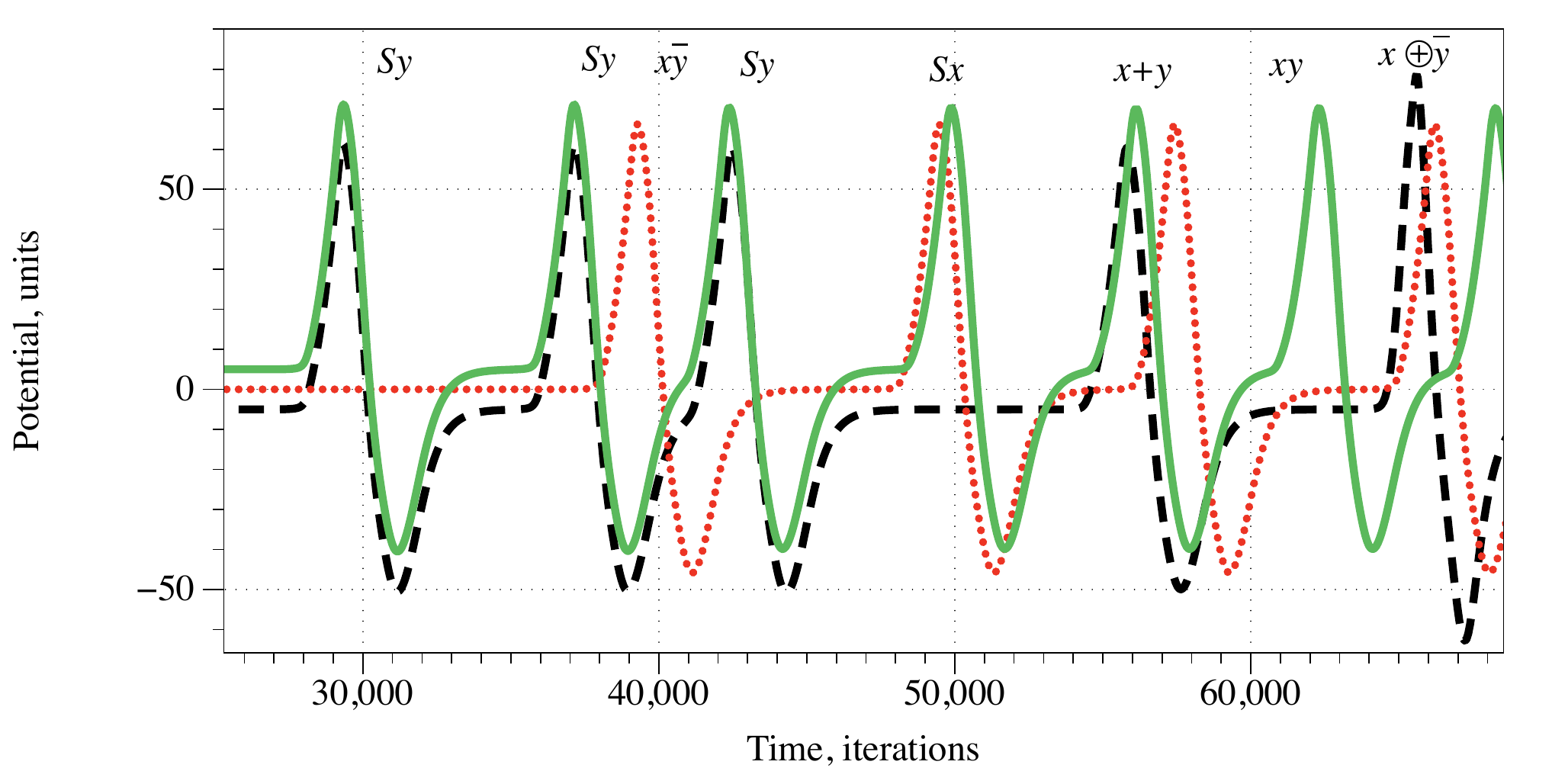}
      \caption{Fragment of electrical potential record on electrode 7 in response to inputs (01), black dashed line, (10), red dotted line, (11), solid green line, entered as impulses via electrodes $E_x=5$ and $E_y=15$. See locations of electrodes in Fig.~\ref{fig:mycelium}d. To make the individual plots visible in places of exact overlapping, we added potential $-5$ to recording in response to input (01) and and potential $5$ to recording in response to input (11). }
    \label{fig:5_15_example}
\end{figure}

\begin{figure}[!tbp]
    \centering
  \includegraphics[height=0.9\textheight]{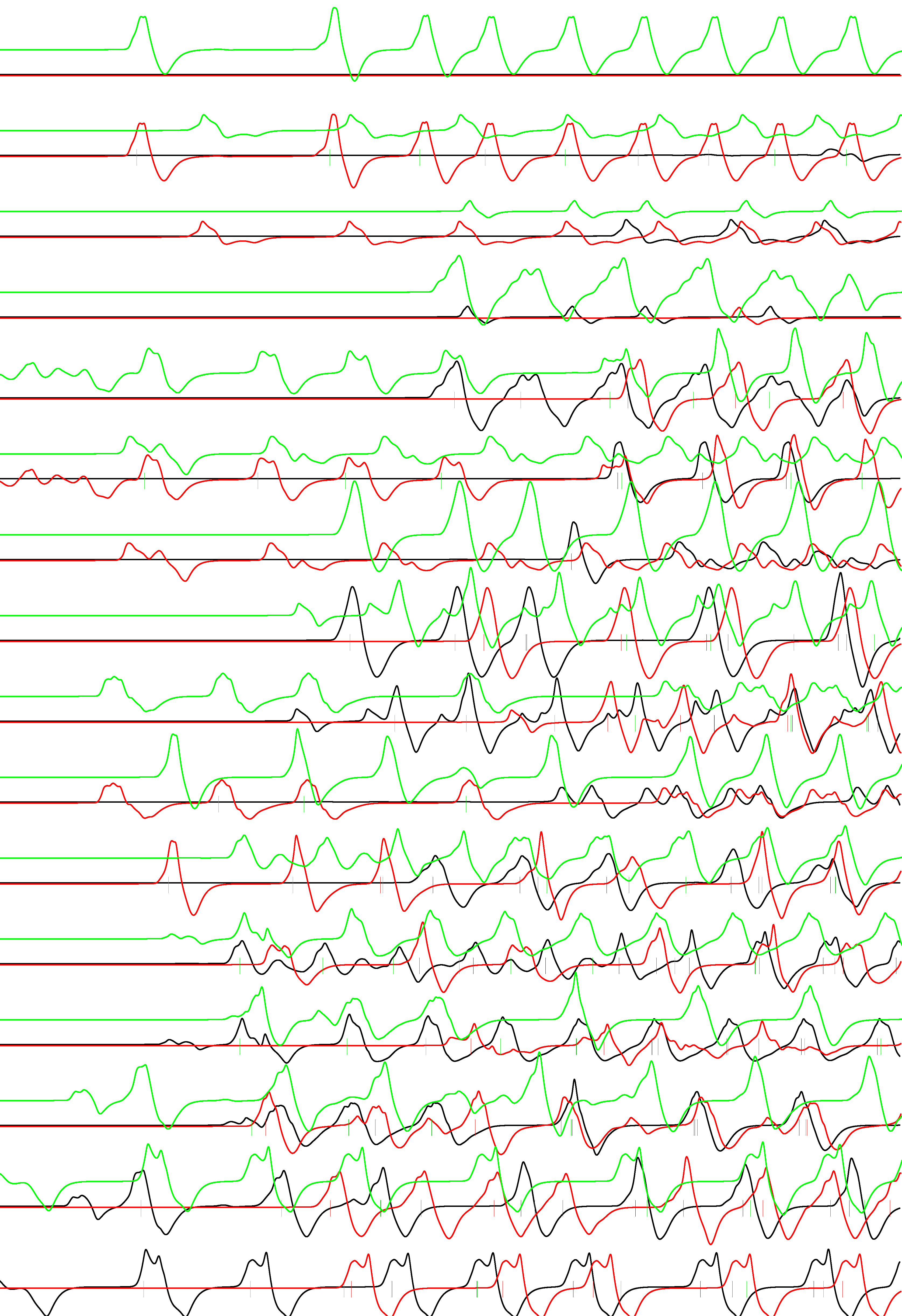}
    \caption{Recording of electrical potential from all electrodes in responses to inputs in response to inputs (01), black  line, (10), red  line, (11), green line, injected as spikes via electrodes $E_x=5$ and $E_y=15$. }
    \label{fig:5_15_allrecordings}
\end{figure}

Input Boolean values are encoded as follows. We earmark two sites of the network as dedicated inputs, $x$ and $y$, and represent logical {\sc True}, or `1', as an excitation, or an impulse injected injected in the network via electrodes. If $x=1$ then the site corresponding to $x$ is excited, if $x=0$ the site is not excited. 

We here assume that each spike represents logical {\sc True} and that spikes occurring within less than $2 \cdot 10^2$ iterations happen {simultaneously}. By selecting specific intervals of recordings we can realise several gates in a single site of recording. In this particular case we assumed that spikes are separated if their occurrences lie more than  $10^3$ iterations apart. An example is shown in Fig.~\ref{fig:5_15_example}.  

\begin{table}[!tbp]
       \subfigure[$E_x=3, E_y=13, c_2=0.095$]{
\tiny{
    \begin{tabular}{c|cccccccc}
$E$	&	$x+y$	&	$Sy$	&	$x\oplus y$	&	$Sx$	&	$\overline{x}y$	&	$x\overline{y}$	&	$xy$	&	Total	\\ \hline
0	&	0	&	0	&	0	&	0	&	0	&	0	&	0	&	0	\\
1	&	0	&	0	&	0	&	2	&	0	&	0	&	0	&	2	\\
2	&	0	&	0	&	0	&	0	&	0	&	0	&	0	&	0	\\
3	&	0	&	0	&	0	&	0	&	0	&	0	&	0	&	0	\\
4	&	1	&	0	&	0	&	7	&	1	&	0	&	0	&	9	\\
5	&	0	&	0	&	0	&	2	&	2	&	0	&	0	&	4	\\
6	&	0	&	0	&	0	&	2	&	0	&	0	&	0	&	2	\\
7	&	1	&	0	&	0	&	8	&	2	&	0	&	0	&	11	\\
8	&	1	&	0	&	0	&	6	&	1	&	0	&	0	&	8	\\
9	&	0	&	0	&	0	&	0	&	1	&	0	&	0	&	1	\\
10	&	0	&	1	&	1	&	2	&	0	&	1	&	2	&	7	\\
11	&	0	&	0	&	0	&	4	&	2	&	0	&	0	&	6	\\
12	&	0	&	0	&	0	&	3	&	2	&	0	&	0	&	5	\\
13	&	1	&	5	&	0	&	0	&	0	&	1	&	0	&	7	\\
14	&	2	&	5	&	0	&	1	&	0	&	1	&	0	&	9	\\
15	&	0	&	1	&	0	&	5	&	2	&	0	&	0	&	8	\\
Total	&	6	&	12	&	1	&	42	&	13	&	3	&	2	&	79	\\
    \end{tabular}
    }
    }
      \subfigure[$E_x=7, E_y=14, c_2=0.095$]{
\tiny{
    \begin{tabular}{c|cccccccc}
$E$	&	$x+y$	&	$Sy$	&	$x\oplus y$	&	$Sx$	&	$\overline{x}y$	&	$x\overline{y}$	&	$xy$	&	Total	\\ \hline
0	&	0	&	0	&	0	&	0	&	0	&	0	&	0	&	0	\\
1	&	0	&	0	&	0	&	5	&	0	&	0	&	0	&	5	\\
2	&	0	&	0	&	0	&	0	&	0	&	0	&	0	&	0	\\
3	&	0	&	0	&	0	&	0	&	0	&	0	&	0	&	0	\\
4	&	0	&	0	&	0	&	6	&	1	&	0	&	0	&	7	\\
5	&	0	&	0	&	0	&	6	&	0	&	0	&	0	&	6	\\
6	&	0	&	0	&	0	&	2	&	0	&	0	&	0	&	2	\\
7	&	1	&	0	&	0	&	4	&	2	&	0	&	0	&	7	\\
8	&	0	&	0	&	0	&	7	&	0	&	0	&	0	&	7	\\
9	&	0	&	0	&	0	&	0	&	0	&	0	&	0	&	0	\\
10	&	1	&	0	&	1	&	5	&	0	&	0	&	0	&	7	\\
11	&	1	&	0	&	0	&	4	&	3	&	0	&	0	&	8	\\
12	&	2	&	0	&	0	&	3	&	0	&	0	&	0	&	5	\\
13	&	1	&	5	&	0	&	1	&	0	&	2	&	0	&	9	\\
14	&	0	&	4	&	0	&	2	&	1	&	2	&	1	&	10	\\
15	&	0	&	0	&	0	&	8	&	2	&	0	&	0	&	10	\\
Total	&	6	&	9	&	1	&	53	&	9	&	4	&	1	&	83	\\
    \end{tabular}
    }
    }
\subfigure[$E_x=7, E_y=14, c_2=0.094$]{
\tiny{
    \begin{tabular}{c|cccccccc}
$E$	&	$x+y$	&	$Sy$	&	$x\oplus y$	&	$Sx$	&	$\overline{x}y$	&	$x\overline{y}$	&	$xy$	&	Total	\\ \hline
0	&	0	&	0	&	0	&	0	&	0	&	0	&	0	&	0	\\
1	&	0	&	0	&	0	&	2	&	1	&	0	&	0	&	3	\\
2	&	0	&	0	&	0	&	0	&	0	&	0	&	0	&	0	\\
3	&	0	&	0	&	0	&	0	&	0	&	0	&	0	&	0	\\
4	&	0	&	0	&	0	&	1	&	3	&	0	&	0	&	4	\\
5	&	0	&	0	&	0	&	1	&	1	&	0	&	0	&	2	\\
6	&	0	&	0	&	0	&	1	&	1	&	0	&	0	&	2	\\
7	&	0	&	0	&	0	&	0	&	2	&	0	&	0	&	2	\\
8	&	0	&	0	&	0	&	3	&	4	&	0	&	0	&	7	\\
9	&	0	&	0	&	0	&	0	&	1	&	2	&	0	&	3	\\
10	&	0	&	0	&	0	&	1	&	1	&	0	&	0	&	2	\\
11	&	0	&	0	&	0	&	0	&	6	&	0	&	0	&	6	\\
12	&	1	&	0	&	1	&	2	&	3	&	1	&	0	&	8	\\
13	&	0	&	2	&	0	&	2	&	0	&	1	&	0	&	5	\\
14	&	1	&	3	&	0	&	0	&	3	&	0	&	2	&	9	\\
15	&	0	&	2	&	0	&	0	&	0	&	1	&	0	&	3	\\
Total	&	2	&	7	&	1	&	13	&	26	&	5	&	2	&	56	\\
    \end{tabular}
    }
    }
    \subfigure[$E_x=5, E_y=15, c_2=0.094$]{
\tiny{
    \begin{tabular}{c|cccccccc}
$E$	&	$x+y$	&	$Sy$	&	$x\oplus y$	&	$Sx$	&	$\overline{x}y$	&	$x\overline{y}$	&	$xy$	&	Total	\\ \hline
0	&	0	&	0	&	0	&	0	&	0	&	0	&	0	&	0	\\
1	&	0	&	0	&	0	&	1	&	1	&	0	&	0	&	2	\\
2	&	0	&	0	&	0	&	0	&	0	&	0	&	0	&	0	\\
3	&	0	&	0	&	0	&	0	&	0	&	0	&	0	&	0	\\
4	&	0	&	1	&	0	&	0	&	2	&	1	&	1	&	5	\\
5	&	0	&	0	&	0	&	0	&	1	&	0	&	0	&	1	\\
6	&	0	&	0	&	0	&	0	&	1	&	0	&	0	&	1	\\
7	&	1	&	2	&	0	&	0	&	0	&	4	&	0	&	7	\\
8	&	0	&	1	&	0	&	1	&	2	&	1	&	0	&	5	\\
9	&	0	&	0	&	0	&	1	&	2	&	0	&	0	&	3	\\
10	&	0	&	0	&	0	&	2	&	0	&	0	&	1	&	3	\\
11	&	1	&	5	&	0	&	0	&	1	&	1	&	1	&	9	\\
12	&	0	&	6	&	0	&	0	&	1	&	2	&	0	&	9	\\
13	&	2	&	0	&	2	&	1	&	1	&	1	&	1	&	8	\\
14	&	0	&	1	&	0	&	1	&	0	&	5	&	0	&	7	\\
15	&	0	&	0	&	0	&	0	&	0	&	1	&	0	&	1	\\
Total	&	4	&	16	&	2	&	7	&	12	&	16	&	4	&	61	\\
    \end{tabular}
    }
    }
    \subfigure[$E_x=5, E_y=15, c_2=0.095$]{
\tiny{
    \begin{tabular}{c|cccccccc}
$E$	&	$x+y$	&	$Sy$	&	$x\oplus y$	&	$Sx$	&	$\overline{x}y$	&	$x\overline{y}$	&	$xy$	&	Total	\\ \hline
0	&	0	&	0	&	0	&	0	&	0	&	0	&	0	&	0	\\
1	&	0	&	0	&	0	&	8	&	0	&	0	&	0	&	8	\\
2	&	0	&	0	&	0	&	0	&	0	&	0	&	0	&	0	\\
3	&	0	&	0	&	0	&	0	&	0	&	0	&	0	&	0	\\
4	&	1	&	4	&	0	&	0	&	0	&	2	&	0	&	7	\\
5	&	3	&	0	&	0	&	4	&	0	&	0	&	0	&	7	\\
6	&	0	&	0	&	0	&	0	&	1	&	0	&	0	&	1	\\
7	&	1	&	3	&	1	&	1	&	0	&	1	&	1	&	8	\\
8	&	0	&	5	&	0	&	1	&	0	&	2	&	0	&	8	\\
9	&	0	&	0	&	0	&	3	&	0	&	0	&	0	&	3	\\
10	&	1	&	0	&	2	&	4	&	2	&	0	&	2	&	11	\\
11	&	2	&	4	&	0	&	2	&	2	&	0	&	1	&	11	\\
12	&	1	&	7	&	0	&	0	&	0	&	3	&	0	&	11	\\
13	&	3	&	1	&	0	&	2	&	0	&	1	&	0	&	7	\\
14	&	1	&	5	&	0	&	0	&	0	&	6	&	0	&	12	\\
15	&	1	&	3	&	1	&	2	&	2	&	1	&	1	&	11	\\
Total	&	14	&	32	&	4	&	27	&	7	&	16	&	5	&	105	\\
    \end{tabular}
    }
    }
    \subfigure[$E_x=5, E_y=15, c_2=0.096$]{
\tiny{
    \begin{tabular}{c|cccccccc}
$E$	&	$x+y$	&	$Sy$	&	$x\oplus y$	&	$Sx$	&	$\overline{x}y$	&	$x\overline{y}$	&	$xy$	&	Total	\\ \hline
0	&	0	&	0	&	0	&	0	&	0	&	0	&	0	&	0	\\
1	&	0	&	0	&	0	&	2	&	1	&	0	&	0	&	3	\\
2	&	0	&	0	&	0	&	0	&	0	&	0	&	0	&	0	\\
3	&	0	&	0	&	0	&	0	&	0	&	0	&	0	&	0	\\
4	&	1	&	3	&	0	&	0	&	0	&	1	&	0	&	5	\\
5	&	0	&	0	&	0	&	4	&	2	&	0	&	0	&	6	\\
6	&	0	&	0	&	0	&	2	&	1	&	0	&	0	&	3	\\
7	&	1	&	4	&	0	&	1	&	0	&	1	&	0	&	7	\\
8	&	1	&	5	&	0	&	0	&	0	&	1	&	0	&	7	\\
9	&	0	&	0	&	0	&	7	&	0	&	0	&	0	&	7	\\
10	&	0	&	1	&	0	&	5	&	1	&	1	&	0	&	8	\\
11	&	2	&	3	&	0	&	1	&	1	&	1	&	1	&	9	\\
12	&	0	&	4	&	0	&	0	&	0	&	0	&	0	&	4	\\
13	&	0	&	2	&	0	&	2	&	0	&	2	&	0	&	6	\\
14	&	1	&	3	&	0	&	2	&	0	&	3	&	0	&	9	\\
15	&	0	&	3	&	0	&	2	&	0	&	2	&	0	&	7	\\
Total	&	6	&	28	&	0	&	28	&	6	&	12	&	1	&	81	\\
    \end{tabular}
    }
    }
    \caption{Numbers of Boolean gates detected for selected pairs of input electrodes $E_x$ and $E_y$.}
    \label{tab:numbers}
\end{table}

\begin{figure}[!tbp]
    \centering
    \includegraphics[scale=0.6]{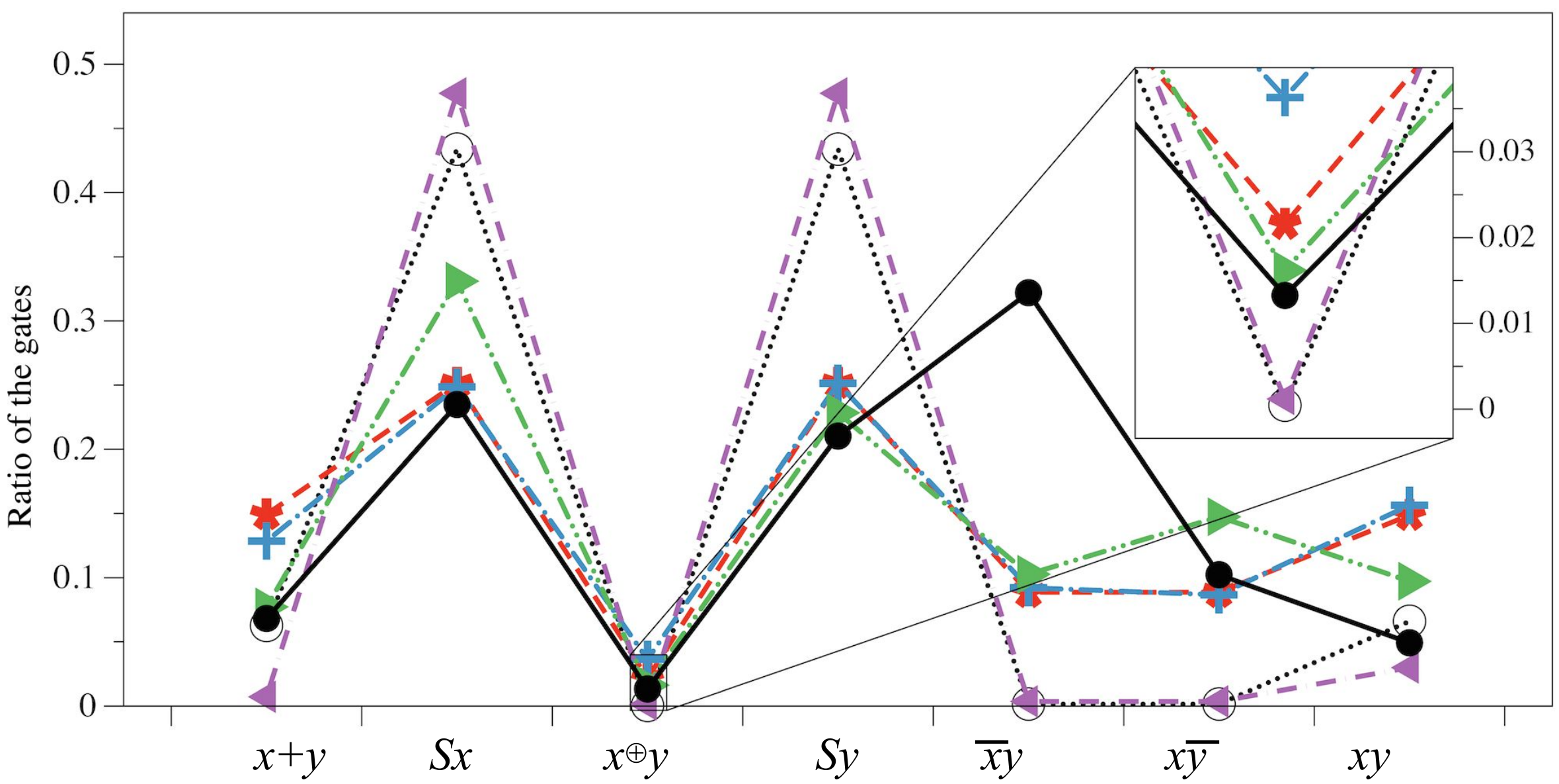}
    \caption{Comparative ratios of Boolean gates discovered in 
    mycelium network in present analysed in present paper, black disc and solid line;
    slime mould \emph{Physarum polycephalum}~\cite{harding2018discovering}, black circle and dotted line;
    succulent plant~\cite{adamatzky2018computers}, red snowflake and dashed line; 
    single molecule of protein verotoxin~\cite{adamatzky2017computing}, light blue `+' and dash-dot line;
    actin bundles network~\cite{adamatzky2019computing}, green triangle pointing right and dash-dot-dot line;
    actin monomer~\cite{adamatzky2017logical}, magenta triangle pointing left and dashed line. Area of {\sc xor} gate is magnified in the insert. Lines are to guide eye only.}
    \label{fig:gatesDistribution}
\end{figure}

Numbers of Boolean gates detected on the electrodes for selected pairs of input electrodes are shown in Tab.~\ref{tab:numbers}. We see that select $x$ and select $y$ gates, $Sx$ and $Sy$ are most frequent. They usually are detected with the same frequency (Tab.~\ref{tab:numbers}d--f), however there are examples of input electrode pairs, where one of the select gates is found much more often  than another. This is most visible for the pair $(E_x,E_y)=(3,13)$ where $Sx$ dominates (Tab.~\ref{tab:numbers}a), and the pair (7,14) and   (Tab.~\ref{tab:numbers}bc) where $Sy$ dominates. Next common gates in the hierarchy are $\overline{x}y$ and $x\overline{y}$. The gates $xy$ and $x+y$ are detected with nearly the same frequency with gate $x+y$ being slightly more common. The $x\oplus y$ is the most rare gate. 

The sub-tables Tab.~\ref{tab:numbers}d, Tab.~\ref{tab:numbers}e and Tab.~\ref{tab:numbers}f show how excitability of the network affects numbers of gates detected. The networks with high excitability, $c_2=0.094$, and low excitability, $c_2=0.0096$ realise smaller number of gates then that realised by sub-excitable network, $c_2=0.095$.

Overall distribution (average of outputs of input electrode pairs (3,13), (5,15), (7,14), (4,13), (13,7)) of a ratio of gates discovered is shown in Fig.~\ref{fig:gatesDistribution}. This is accompanied by distributions of gates discovered in experimental laboratory reservoir computing with slime mould \emph{Physarum polycephalum}~\cite{harding2018discovering}, succulent plant~\cite{adamatzky2018computers} and numerical modelling experiments on computing with 
protein verotoxin~\cite{adamatzky2017computing},    actin bundles network~\cite{adamatzky2019computing}, and actin monomer~\cite{adamatzky2017logical}. All the listed distributions have very similar structure with gates selecting one of the inputs in majority, followed by {\sc or} gate, {\sc not-and} an {\sc and-not} gates. The gate {\sc and} is usually underrepresented in experimental and modelling experiments. The gate {\sc xor} is a rare find.

\section{Discussion}
\label{discussion}

We have demonstrated how sets of logical gates can be implemented in single colony mycelium networks via initiation of electrical impulses. The impulses travel in the network, interact with each other (annihilate, reflect, change their phase). Thus for different combinations of input impulses and record different combinations of output impulses, which in some cases can be interpreted as representing two-inputs-one-output functions. 

To estimate a speed of computation we refer to Olsson and Hansson's~\cite{olsson1995action} original study, in which they proposed that electrical activity in fungi could be used for communication with message propagation speed 0.5~mm/sec. Diameter of the colony (Fig.~\ref{fig:mycelium}a), which experimental laboratory images has been used to run FHN model, is c.~1.7~mm. Thus, it takes the excitation waves initiated at a boundary of the colony up to 3-4~sec to span the whole mycelium network (this time is equivalent to c. 70K iterations of the numerical integration model). In 3-4~sec the mycelium network can compute up to a hundred logical gates. This gives us the rate of a  gate per 0.03~sec, or, in terms of frequency this will be c.~30~Hz. The mycelium network computing can not compete with existing silicon architecture however its application domain can be a unique of living biosensors (a distribution of gates realised might be affected by environmental conditions)~\cite{manzella2013plants} and computation embedded into  structural elements where fungal materials are used~\cite{ross2016your,heisel2018design,dahmen2016soft}. 

\section{Acknowledgement}

This project has received funding from the European Union's Horizon 2020 research and innovation programme FET OPEN ``Challenging current thinking'' under grant agreement No 858132.


\end{document}